\documentclass[arxiv]{myproc}  

\usepackage{macros}

\title{Buying Data of Unknown Quality: Statistical Information Procurement Auctions}

\author{Yuchen Hu$^{1}$, Martin J. Wainwright$^{2}$, Stephen Bates$^{2}$\\[0.5em]
$^{1}$Columbia University \\
$^{2}$MIT}

\date{ \today}

\begin{document}

\myprocmaketitle

\begin{abstract}
We study statistical parameter estimation in the setting of data
markets when providers differ in both provision costs and estimation-relevant data quality.
We define
a cost-per-information score that summarizes each provider’s provision
cost per unit of information about the buyer’s estimation objective.
When quality is known ex ante, we describe a second-score procurement mechanism that ranks providers by
this score, and endogenously chooses both a provider and a sample size
while making truthful cost reports optimal. We then turn to the more
realistic setting where data quality is private, and can only be assessed noisily via the delivered data. In this setting, we
propose a simple mechanism that augments the second-score rule with a
lenient ex post statistical test of the reported quality. 
We prove that, under mild conditions, the mechanism admits near-truthful reports that become approximately optimal and individually rational as the procured sample size grows. Under these near-truthful reports, the buyer asymptotically recovers the performance of the corresponding known-quality second-score mechanism.
Our analysis highlights
how the choice of verification test and the buyer’s accuracy-cost
tradeoff jointly shape participation and misreporting incentives in
data markets.
\end{abstract}

\section{Introduction}

Many organizations now purchase data at scale for statistical estimation and machine-learning tasks, either by contracting with external providers or by outsourcing parts of their data collection to specialized teams \citep{liang2018survey}.  In these settings, data are valuable because they are informative about a target parameter: some datasets are cheap but noisy or weakly informative, while others are expensive because they require careful measurement, recruiting hard-to-reach populations, or running complex experiments~\citep{haug2011costs}.
Yet, before any contract is signed, it is difficult to know how useful a provider’s data will be for the buyer’s estimation objective, and even harder to verify the provider’s internal cost. This information asymmetry raises a central question for data procurement: \emph{How can a buyer elicit sellers’ private information while deciding from whom to buy, how much data to acquire, and how much to pay?}

Classical second-price procurement auctions address one part of this problem: when each seller has a single private cost parameter, they make truthful cost reporting a weakly dominant strategy and select the cheapest provider~\citep{vickrey1961counterspeculation,myerson1981optimal}. 
However,
applying this logic to data procurement is more subtle because the good being purchased is statistical information. 
The buyer must choose both a seller and a sample size, and the optimal quantity depends jointly on cost and on how informative the seller’s data are for the buyer’s target estimation task. 
Moreover, each seller’s data quality is typically hidden before purchase, which renders the auction problem multi-dimensional. It is
thus in general difficult to enforce truthful reporting via a single
scalar bid, as sellers can adjust the different components of their
report to alter the purchased quantity (and thus their profit) without
necessarily changing the
bid~\citep{armstrong1996multiproduct,rochet1998ironing}.


\subsection{Our contributions}

Our contributions are threefold. First, we formulate a data procurement problem in which a buyer trades off statistical accuracy against procurement cost when sellers differ in both provision costs and data quality. We propose a second-score procurement mechanism based on a “cost-per-information” score. When data quality is known, this mechanism selects the most cost-effective seller, chooses the sample size endogenously, and makes truthful cost reporting optimal.

Second, we study the realistic case where data quality is private and can only be assessed noisily from the purchased data. We augment the score-based procurement rule with an ex post statistical verification step that checks the winning seller’s reported quality using delivered data. This preserves the simple allocation and payment structure above, while discouraging large quality misreports. 

Third, we characterize the reporting incentives induced by noisy statistical verification. Under mild concentration conditions on the verification statistic, we show that every interior seller has a near-truthful report---a slightly conservative quality claim together with a small price markup---that is approximately optimal against any realized profile of the other sellers’ reports and yields nonnegative expected utility. Both distortions vanish as the procured sample size grows. These pointwise incentive guarantees induce a near-truthful approximate Bayesian Nash equilibrium, under which the buyer incurs only a vanishing additional loss relative to the corresponding known-quality second-score mechanism.

\begin{figure}[t]
\centering
\begin{minipage}[c]{0.495\linewidth}
\centering
\resizebox{\linewidth}{!}{
\begin{tikzpicture}
  \node[box, minimum width=2.55cm] (type) at (0,0) {
    Seller $i$\\[-1pt]
    private type $(c_i,\FishInv_i)$
  };
  \node[box, minimum width=2.92cm, right=1.42cm of type] (report) {
    Report\\[-1pt]
    $(p_i,\FishInvTil_i)$
  };
  \draw[bluearrow] (type) -- node[above, label] {strategic\\claim} (report);

  \node[box, minimum width=3.95cm, below=0.76cm of report] (score) {
    Score reports; select winner\\[-1pt]
    set payment and sample size
  };
  \draw[arrow] (report) -- (score);

  \node[box, minimum width=3.70cm, below=0.70cm of score] (verify) {
    Statistical verification\\[-1pt]
    using delivered data
  };
  \draw[greenarrow] (score) -- node[right=1pt, label] {winner\\delivers data} (verify);

  \node[result, minimum width=1.72cm, below left=0.62cm and 0.10cm of verify] (pass) {
    Pass\\[-1pt]
    pay seller
  };
  \node[result, minimum width=1.85cm, below right=0.62cm and 0.10cm of verify] (fail) {
    Fail\\[-1pt]
    void contract
  };
  \coordinate (split) at ($(verify.south)+(0,-0.24)$);
  \draw[greenarrow] (verify.south) -- (split) -| (pass.north);
  \draw[arrow] (split) -| (fail.north);

  \node[fit=(type)(report)(score)(verify)(pass)(fail), inner sep=0pt] (leftpanel) {};
\end{tikzpicture}}
\end{minipage}
\hfill
\begin{minipage}[c]{0.495\linewidth}
\centering
\includegraphics[width=\linewidth]{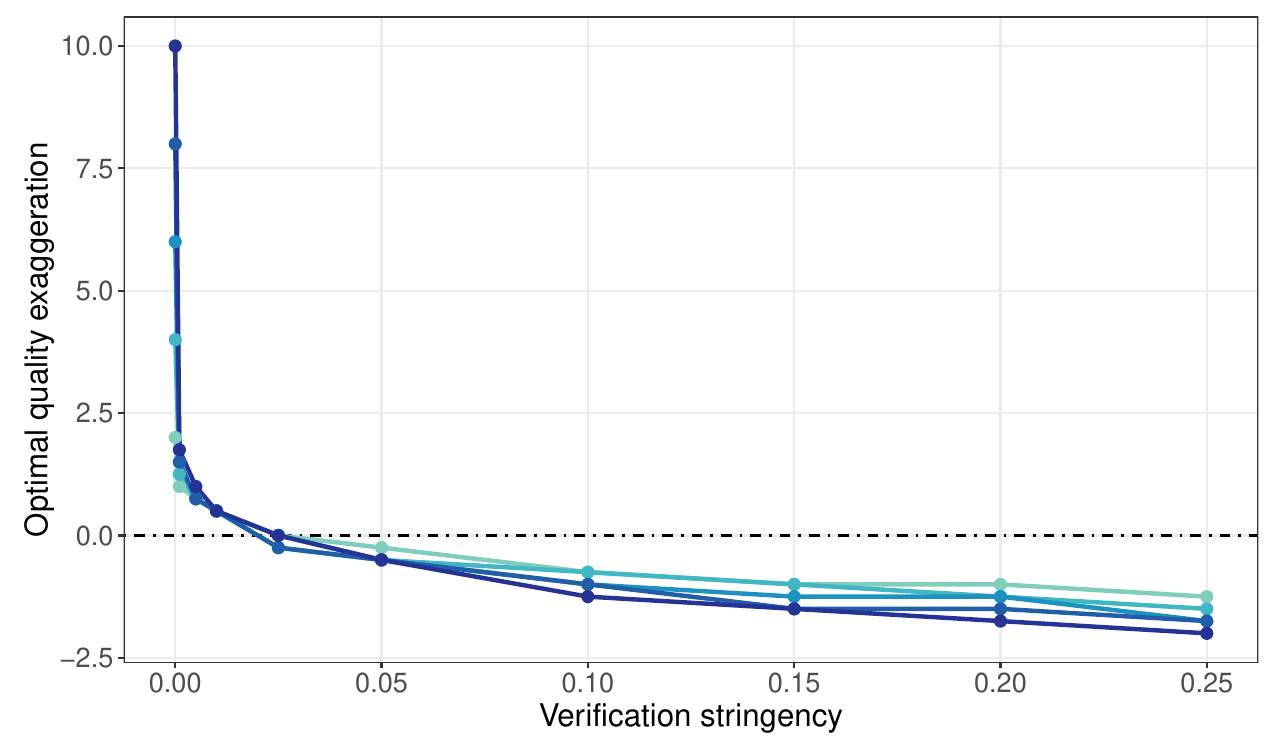}
\end{minipage}
\caption{Statistical verification discourages quality exaggeration. Left: sellers report price and claimed quality; the mechanism selects a winner, sets payment and sample size, and verifies claimed quality using delivered data. Right: Stronger verification reduces optimal quality exaggeration; darker curves correspond to worse true quality.}
\label{fig:verification_mechanism}
\end{figure}

\Cref{fig:verification_mechanism} gives a preview of our mechanism. Without verification, sellers can have strong incentives to exaggerate quality. Once the mechanism subjects the winning seller’s claim to even a mild statistical check, such exaggeration becomes much less attractive. Our results provide a theoretical foundation and offer guidance on the choice of sample size and verification tests. Overall, our mechanism could improve transparency in data markets by giving buyers a way to discipline exaggerated quality claims.

\subsection{Related work}
\label{SecRelated}

Our work builds on the classic literature in auction theory and
mechanism design~\citep{laffont2002theory}. Truthful mechanism design
originates from the Vickrey--Clarke--Groves (VCG)
framework~\citep{vickrey1961counterspeculation,clarke1971multipart,groves1973incentives},
which characterizes efficient and incentive-compatible mechanisms
under quasi-linear preferences. \cite{myerson1981optimal} further
provides a general theory of optimal auctions in single-parameter
environments that maximize the seller’s expected revenue under
incentive and participation constraints.  These frameworks have since
inspired a broad class of contracting mechanisms that aim to elicit
truthful cost reports while maximizing allocation efficiency or
minimizing expenditure.

The procurement setting we study is a natural dual of the auction
problem, where a buyer seeks to minimize procurement costs while
ensuring sufficient quality of the procured goods or
services~\citep{laffont1993theory}.  In these settings, the buyer
faces the reverse problem of Myerson’s auction: rather than allocating
goods to bidders with the highest virtual values, the principal
selects suppliers with the lowest virtual costs.  In procurement
auctions, the principal often needs to balance price and quality
considerations~\citep{klein1981role,manelli1995optimal,asker2010procurement,yao2020price}.
A key challenge arises when the buyer does not know the supplier’s
quality ex ante.  Prior work has addressed this by modeling
procurement with costly pre-contract qualification of
suppliers~\citep{wan2009rfq,chen2018optimal}; by studying mechanisms
with exact ex post
verification~\citep{ben2014optimal,mylovanov2017optimal};
and by leveraging
relational and reputational enforcement, such as contract termination
or discriminatory scoring mechanisms in repeated
relationships~\citep{board2011relational,auriol2017economic,albano2017public}.
Our study contributes to this line of research by proposing a
procurement mechanism with statistical quality verification, where the
buyer cross-checks the reported data quality using noisy empirical
information estimates and voids the contract when the verification
fails.

Perhaps most closely related to our work is the growing literature on
data procurement for improving the performance of statistical
estimation and machine learning models.  Like this work, many of these
studies build upon Myerson’s seminal framework and adapt it to address
challenges unique to data acquisition.  One prevalent design for data
procurement with multiple agents is to use peer prediction (see,
e.g.,~\citealt{miller2005eliciting,richardson2019rewarding,chen2020truthful,clinton2025cram}
and references therein), which leverages mutual information between
reports to make truth-telling an equilibrium.  Beyond static designs,
recent work has studied data procurement mechanisms in online settings,
where the principal sequentially acquires data while balancing
exploration, budget constraints, and incentive compatibility
\citep{abernethy2015low,cao2017data,an2017towards,chen2018optimal,chen2019prior,li2021data}.
Our work is also closely related to research on data collection
delegation
\citep{cai2015optimum,gast2020linear,ananthakrishnan2024delegating, saig2023delegated},
in which data providers act as “workers” who can improve the quality
of their data through costly effort, and mechanisms are designed to
incentivize such effort so that high-quality data are produced at low
cost for statistical estimation. In contrast, we consider a setting
where the quality of each agent’s data is fixed and exogenous, and the
principal’s goal is to select agents with inherently higher data
quality rather than incentivizing them to improve it.

The mechanism we study relies on a verification test of the amount of
statistical information contained in the reported data, compared to
the information implied by the agent’s report. In this spirit, our
work is related to information elicitation.  For prediction and estimation, proper
scoring rules have been widely used to elicit truthful reports of
probabilistic forecasts and sufficient
statistics \citep{gneiting2007strictly,lambert2008eliciting,frankel2019quantifying}.
For decision making, they have been extended to design mechanisms that
align incentives with optimal actions or posterior beliefs in
decision-theoretic
settings \citep{osband1989optimal,chen2011information,shi2024sharp,shi2025instance},
and, in particular, to develop decision scoring rules that incentivize
agents to truthfully report costly
information \citep{oesterheld2020minimum,neyman2021binary,li2022optimization,papireddygari2022contracts}. Compared
with these works, we study a more challenging setting in which agents
may benefit from misreporting the information required for the
principal’s decision by an unknown and unobservable amount, which we
mitigate through an auction-based design.

Finally, at a high level, our work contributes to the growing
literature on mechanism design for data sharing and data markets
\citep{liang2018survey,castro2023data}. Much of this literature
focuses on developing fair and truth-incentivizing data valuation
strategies implemented through centralized data-access platforms or
third-party intermediaries
\citep{jia2019efficient,ghorbani2019data,yoon2020data,wang2023data,jiang2023opendataval,just2023lava,zhao2023addressing,huang2023evaluating,lu2024daved},
as well as on privacy-preserving mechanisms for data procurement,
exchange, and
aggregation~\citep{ghosh2011selling,fleischer2012approximately,li2014theory,anjarlekar2023striking,cummings2023optimal,fallah2024optimal}. In
contrast, we study a decentralized data procurement problem, where
data remain privately held by self-interested agents, and the
principal must design incentives to elicit and verify data quality
without relying on a trusted intermediary.


\section{Problem set-up}
\label{SecSetup}

In this section, we begin by setting up the problem more
precisely, including the goals of the statistician
(\Cref{SecStatGoal}) and the generic buyer-seller interaction
(\Cref{SecInter}) that underlies the class of procurement auctions
that we study.

\subsection{Statistical Goal}
\label{SecStatGoal}

A data buyer, also referred to as the principal, wishes to estimate
a target parameter $\theta \in \Theta$ using data. 
Each provider $\agentindex \in
[\numagent] \defn \{1, \ldots, \numagent \}$ can generate samples relevant to this estimation objective, but providers may differ in how informative their samples are. We summarize this task-relevant informativeness by a scalar $\InvFish_\agentindex > 0$, where smaller values of $\InvFish_\agentindex$ correspond to more informative data. If the buyer purchases $\numobs_\agentindex$ samples from provider $\agentindex$, we assume that the resulting \emph{estimation error} scales as
\begin{align}
\label{eq:mse_form}
\operatorname{MSE} \asymp 
\p{\InvFish_\agentindex/\numobs_\agentindex}^\rho,
\end{align}
The exponent $\rho\in(0,1]$ captures the \emph{sample-complexity} of the underlying statistical problem.
The usual parametric rate corresponds to $\rho=1$, while values $\rho<1$ capture slower rates that arise in nonparametric estimation problems.
For most of the main results, we focus on the parametric case $\rho=1$ to keep notation simple. The same mechanism extends to the general scaling $\rho<1$; we state this extension in~\Cref{sec:sub_loss}. 

Under the parametric scaling $\rho=1$, $\InvFish_\agentindex$ can be interpreted as \emph{inverse Fisher information}. If provider $\agentindex$ supplies observations with \emph{Fisher information} $\Fish_\agentindex > 0$ for $\theta$, then $\InvFish_\agentindex = 1/\Fish_\agentindex$, and purchasing $\numobs_\agentindex$ samples yields error of order $1/(\numobs_\agentindex
  \Fish_\agentindex)=\InvFish_\agentindex/\numobs_\agentindex$.
In a Gaussian location model where provider $\agentindex$ generates samples from the normal distribution
$N(\theta, \sigma_\agentindex^2)$, we have $\InvFish_\agentindex \equiv \sigma_\agentindex^2$. More generally, in regular parametric models, $\InvFish_\agentindex$ can be taken as inverse Fisher information, or as an appropriate scalar summary of the inverse information matrix when $\theta$ is vector-valued.


\subsection{Buyer-Seller Interaction}
\label{SecInter}

In a data procurement auction, the data buyer
receives a price $\price_\agentindex$ from each agent $\agentindex \in
\optinset$ in the subset $\optinset \subseteq [\numagent]$ of agents
who chose to opt-in, referred to as the \emph{opt-in set}.  The
statistician seeks to balance their procurement cost with statistical
accuracy; this tradeoff is controlled by a pre-specified scalar $\beta
> 0$. If the statistician buys $\numsamp_\agentindex$ samples at price
$\price_\agentindex$ from agent $\agentindex$, then their
corresponding loss is given by the weighted sum
\begin{align}
\label{eq:principal_loss}
  \Loss(\agentindex, \price_{\agentindex},\numsamp_{\agentindex}) &
  \defn \beta \cdot
  \underbrace{\p{\FishInv_{\agentindex}/\numsamp_{\agentindex}}^\rho}_{\mbox{Error}}
  + \underbrace{\price_{\agentindex}
    \numsamp_{\agentindex}}_{\mbox{Data cost}},
\end{align}
where we refer to $\FishInv_\agentindex$ as the (inverse) \emph{data quality} that controls the estimation error.
Trading off statistical accuracy against data-acquisition cost is standard in optimal experimental design and strategic data acquisition. While richer statistical problems may require multidimensional notions of quality, the scalar-quality formulation provides a simple and interpretable benchmark for isolating the challenge created by privately known data quality.

The principal wishes to select a seller
$\optagent \in \optinset$ and a sample size
$\numsamp_{\optagent}$ to purchase so as to obtain the minimal loss. We focus on single-provider procurement because, in our scalar-information setting, the buyer’s optimal allocation purchases all samples from the provider with the lowest cost per unit of information.

On the seller (or data provider) side, there are costs associated
with generating and providing data.  More precisely, seller
$\agentindex$ incurs cost $\cost_\agentindex$ to produce a sample;
this per-sample cost can reflect privacy concerns, computational
burden, or other opportunity costs.  Thus, seller $\agentindex$'s
private information can be summarized by the type pair
$\type_\agentindex \defn (\cost_\agentindex, \FishInv_\agentindex)$. Throughout, we treat $\rho$ as known and common across providers, since it is determined by the buyer's estimation problem.

Suppose that the procurement auction leads to seller $\optagent$ being
selected. In this case, seller $\optagent$ receives payment
$\PurchPrice \numsamp_{\optagent}$ and incurs cost
$\cost_{\optagent} \numsamp_{\optagent}$, so that their net utility is
given by the difference $(\PurchPrice - \cost_{\optagent})
\numsamp_{\optagent}$. Sellers who are not selected receive zero
utility. We are interested in designing procurement mechanisms that
elicit information about sellers' types and jointly determine the
triple $(\optindex, \numsamp_\optindex, \price_\optindex)$, subject to
incentive and participation constraints.


\section{Main Results}

In this section, we develop our information procurement mechanisms and study their theoretical properties.  We start by analyzing an idealized benchmark
in which the principal knows each seller's data quality \emph{ex
ante}.  In~\Cref{sec:mech_truth}, we introduce a simple
second-price-per-information mechanism,
Mechanism~\ref{mech:cost_only}, and show that it is truthful and
individually rational. We then turn to the realistic case in which
data quality is privately known to the sellers and can only be
assessed \emph{ex post}.  In~\Cref{sec:mech_noisy}, we introduce an augmented
second-price-per-information mechanism,
Mechanism~\ref{mech:cost_and_quality}, together with a general class
of verification tests. 
Our main results in
\Cref{sec:incentive_analysis} establish that (i) every interior seller has a near-truthful report that is approximately optimal against any realized profile of the other sellers’ reports and yields nonnegative expected utility (\Cref{theorem:incentive}); (ii) these reports induce a near-truthful approximate Bayesian Nash equilibrium (\Cref{coro:BNE}); and (iii) under this equilibrium, the principal’s regret consists of the usual second-score competition gap plus an additional distortion from noisy verification that vanishes in the large-sample limit (\Cref{proposition:principal_loss}).

\subsection{Truthful Mechanism with Known Quality}
\label{sec:mech_truth}

Recall that each agent $\agentindex$ is associated with the type
$\type_\agentindex \defn (\cost_\agentindex, \FishInv_\agentindex)$ of
(inverse) data quality $\FishInv_\agentindex$ and data cost
$\cost_\agentindex$.  In this section, we begin by studying the
simpler setting in which the principal is given \emph{a priori} knowledge
of the data qualities $\{\FishInv_\agentindex
\}_{\agentindex=1}^\numagent$ across the full agent population.  The
per-sample costs $\{\cost_\agentindex \}_{\agentindex=1}^\numagent$
remain private.  We design a simple one-shot mechanism that elicits
cost bids, is truthful and individually rational, and yields a
transparent characterization of the principal’s loss.

Note that our problem can be viewed as an instance of quality-adjusted
procurement: the principal cares about both the cost of acquiring data
and the statistical precision it delivers. Our protocol is based upon
the score $\score_\agentindex \defn \price_\agentindex
\FishInv_\agentindex \equiv \price_\agentindex/\Fish_\agentindex$,
corresponding to the \emph{price per unit Fisher information} for
buying from agent $\agentindex$.  Thus, this score can be regarded as
a cost-effectiveness measure for learning precision.  We note that
similar ``score-based'' procurement rules are widely used in
practice. Examples include public agencies evaluating bids based on a
weighted combination of price and quality, or firms selecting vendors
using a cost-per-performance index~\citep{laffont1993theory}. Our
mechanism formalizes this idea for data acquisition, with statistical
precision playing the role of quality.

\mygraybox{
  \begin{mech}
Second-Price-per-Information Mechanism.
\begin{enumerate}
\item \textbf{Bidding}: Each agent $\agentindex \in [\numagent]$
  decides whether to enter the mechanism, leading the \emph{opt-in
  set} $\optinset \subseteq [\numagent]$ of participating agents. Each
  opt-in agent $\agentindex \in\optinset$ submits a per-sample price
  bid $\price_\agentindex$.
\item \textbf{Scoring}: The principal computes the scores
  $\score_\agentindex \defn \price_\agentindex \FishInv_\agentindex$
  for each $\agentindex \in \optinset$.
  \item \textbf{Selection}: The principal chooses the agent $\optagent
    \defn \arg \min \limits_{\agentindex \in \optinset}
    \score_\agentindex$ with the lowest score.\footnote{Any ties are
    broken according to a fixed rule.}
\item \textbf{Unit Payment}: Letting $s_{(2)} \defn \min
  \limits_{\agentindex\in \optinset \backslash \{\optindex \}}
  \score_\agentindex$ denote the smallest\footnote{If only one seller
  participates (i.e., $|\optinset|=1$), we define $s_{(2)}$ as a fixed
  upper bound on the scores.}  score among the losing agents, the
  winner $\optagent$ is paid a per-sample price
\begin{subequations}
\begin{align}
\PurchPrice \defn \frac{s_{(2)}}{\FishInv_{\optagent}} \; = \; s_{(2)} \Fish_{\optagent}
\end{align}
corresponding to the second-best score scaled by the winner's Fisher information.
\item \textbf{Quantity}: The principal purchases a total of
\begin{align}
\numsamp_{\optagent} = \frac{\sqrt{\beta} \ \FishInv_{\optagent}}{\sqrt{
    s_{(2)}}} \qquad \text{samples from seller $\optagent$.}
\end{align}
\end{subequations}
\end{enumerate}
\label{mech:cost_only}
\end{mech}
}
\noindent Note that conditional on the competitors' bids, the
principal purchases fixed total amount of information from the winner,
regardless of the winner’s own report as long as they win.\\

\noindent Mechanism~\ref{mech:cost_only} can be viewed as a
second-score procurement mechanism with an endogenous purchasing
quantity. Let $s_{(1)}$ and $s_{(2)}$ denote the lowest and
second-lowest scores among the participating agents; by construction,
the winner has score $s_{(1)}$ and is paid based on score $s_{(2)}$.
In particular, the winning agent is paid an information-adjusted
per-sample price $\PurchPrice$ that yields the same price per unit of
information as buying from the runner-up.  Furthermore, recall from
the principal's loss function~\eqref{eq:principal_loss} that, for a
fixed per-sample price $p_\agentindex$, the optimal sample size
$\numsamp_\agentindex^*$ to purchase follows from the first-order
optimality condition for the loss function~\eqref{eq:principal_loss}.
In particular, we have
\begin{equation}
    \frac{\partial\Loss}{\partial \numsamp_\agentindex}(\agentindex,
    \price_{\agentindex},\numsamp_{\agentindex}) \Big|_{\numsamp_\agentindex = \numsamp^*_\agentindex} \; = \; -\beta
    \frac{\FishInv_\agentindex}{(\numsamp_\agentindex^*)^2} +
    \price_\agentindex = 0 \qquad \Rightarrow\qquad
    \numsamp_\agentindex^* = \sqrt{\frac{\beta
        \FishInv_\agentindex}{p_\agentindex}}.
\label{eq:opt_allocation}
\end{equation}
This square-root rule is analogous to Neyman allocation in stratified sampling and to A-optimal design for one-parameter linear Gaussian models, where optimal sample sizes are proportional to the standard deviation divided by the square root of the per-unit cost (see, e.g., \citealt{neyman1934two}, \citealt{pukelsheim2006optimal}).
Now, in the mechanism, the principal uses the information-adjusted second price $\PurchPrice$, which, when substituted into~\eqref{eq:opt_allocation}, gives exactly the sample size $\numsamp_\optagent$ in Mechanism~\ref{mech:cost_only}.
Together, these choices ensure that the
winning agent’s payment does not depend on their own report, and thus
makes the mechanism truthful and individually rational.  

Under this
mechanism, the principal’s total loss is given by
\begin{align*}
\Loss(\optagent, \PurchPrice, \numsamp_{\optagent}) = \beta
\frac{\FishInv_{\optagent}}{\numsamp_{\optagent}} +
\PurchPrice\numsamp_{\optagent} = 2\sqrt{\beta \;
  \score_{(2)}},
\end{align*}
which coincides with the loss from optimally buying from the runner-up
at their true cost under full information.  Therefore, the principal
incurs a $2 \sqrt{\beta} (\sqrt{\score_{(2)}}
-\sqrt{\score_{(1)}})$-gap from the first-best solution.  This
structure is analogous to standard second-price and scoring auctions,
where truthful implementation replaces the best supplier’s terms by
those of the marginal competitor, so that efficiency loss is driven
entirely by the gap between the first and second-best
scores~\citep{vickrey1961counterspeculation,myerson1981optimal}.  The
structure of Mechanism~\ref{mech:cost_only} is robust to more general
scalings of the mean-squared error (MSE).  In~\Cref{sec:sub_loss}, we
describe an extension (Mechanism~\ref{mech:cost_only_sub}) that
applies when the MSE decays at a slower rate, as is expected for
non-parametric estimation problems.

Having set up this mechanism, let us now formalize its key property:
it is truthful and individually rational.
\begin{proposition}
\label{proposition:DSIC_cost}    
Consider a collection of $\numagent$ agents, where agent $i$ has
private cost $\cost_\agentindex$ and publically known data quality
$\FishInv_\agentindex$.  Then truthfully bidding $\price_\agentindex =
\cost_\agentindex$ is a weakly dominant strategy under
Mechanism~\ref{mech:cost_only}.
\end{proposition}
\noindent See~\Cref{sec:proof_DSIC_cost} for the proof.  In this
particular case, with a single unknown parameter $\cost_i$ per agent,
the argument follows along the lines of the VCG
logic~\citep{vickrey1961counterspeculation,clarke1971multipart,groves1973incentives}
for single-parameter environments: Mechanism~\ref{mech:cost_only} is a
second-price auction for units of information, so truthful cost
reporting is weakly dominant.

The assumption of known data quality $\FishInv_\agentindex$ is natural
in certain settings, including when the principal can benchmark
potential suppliers on public test tasks, or data quality is certified
by regulation. However, there exist many data markets in which data
quality remains private to the seller.  When the data qualities
$\FishInv_\agentindex$ are no longer known and must instead be
reported by the agents, Mechanism~\ref{mech:cost_only} no longer
ensures truthfulness.  Suppose that agent $\agentindex$ reports
inverse data quality $\FishInvTil_\agentindex$.  Since the resulting
score $\score_\agentindex = \price_\agentindex
\FishInvTil_\agentindex$ is strictly increasing in
$\FishInvTil_\agentindex$, the agent has incentive to report a lower
value.  Furthermore, conditional on winning, the total payment is
independent of the report, while the quantity supplied decreases with
$\FishInvTil_{\agentindex}$. Hence, by reporting a smaller
$\FishInvTil_\agentindex$, the seller can always increase their
profit.  This difficulty is not specific to our design: when bidders
have multi-dimensional private information, extending Vickrey’s
second-price framework to ``second-score'' auctions is in general
challenging, as a single scalar score typically cannot capture all the
relevant incentive structures in the
problem~\citep{armstrong1996multiproduct,rochet1998ironing}.


\subsection{Second-Score Mechanism with Quality Verification}
\label{sec:mech_noisy}

As previously discussed, when the data quality is private, agents
benefit from under-reporting their (inverse) data quality
$\FishInv_\agentindex$---that is, by inflating the statistical utility
of the data.  In this section, we introduce a verification-based
mechanism designed to mitigate this incentive.

The mechanism relies on an empirical test of Fisher information; it
penalizes agents for mis-reporting by voiding the contract if the
delivered data exhibit a higher inverse Fisher information than
reported.  As we show in the sequel, although this mechanism does not
always guarantee exact truthfulness, it makes substantial
mis-reporting unprofitable with high probability, thereby achieving
approximate or ``almost truthful'' behavior.

The following mechanism extends Mechanism~\ref{mech:cost_only} to the
setting where data qualities are reported by agents and verified
\emph{ex post} through a quality test:
\mygraybox{
\begin{mech}
Second-Price-per-Information Mechanism with Statistical Verification.
\begin{enumerate}
\item \textbf{Bidding}: Each agent $\agentindex$ in the opt-in set
  $\optinset$ reports the pair $(\price_\agentindex,
  \FishInvTil_\agentindex)$.
\item \textbf{Scoring}: The principal computes the scores
  $\score_\agentindex \defn \price_\agentindex
  \FishInvTil_\agentindex$ for each opt-in agent $\agentindex \in
  \optinset$.
\item \textbf{Selection}: The principal selects the agent $\optagent =
  \arg \min \limits_{\agentindex \in \optinset} \score_\agentindex$.
\item \textbf{Unit Payment}: Letting $s_{(2)} \defn \min
  \limits_{\agentindex \in \optinset \backslash \{\optagent \}}
  \score_\agentindex$ denote the smallest score among the losing
  agents, the winner’s per-sample payment is
\begin{subequations}
  \begin{align}
\PurchPrice & \defn
\frac{\score_{(2)}}{\FishInvTil_{\optagent}}.
\end{align}
\item \textbf{Quantity}: The principal purchases
\begin{align}
\numsamp_{\optagent} \defn \frac{\sqrt{\beta} \;
  \FishInvTil_{\optagent}}{\sqrt{\score_{(2)}}}
\end{align}
\end{subequations}
samples from the winning agent $\optagent$.
\item \textbf{Verification Test}: After collecting the data, the
  principal computes an estimate $\FishInvHat_{\optagent}$ of the
  inverse Fisher information based on $\numsamp_{\optagent}$
  samples. The contract is voided\footnote{If the contract is voided,
  the principal pays nothing while the agent still incurs the data
  provision cost.}  if $\FishInvHat_{\optagent} >
  \FishInvTil_{\optagent}$.  
\end{enumerate}
\label{mech:cost_and_quality}
\end{mech}
}

In terms of the report $\FishInvTil_\optagent$ of the winning agent,
the principal's loss is given by
\begin{align}
\Loss(\optagent,\PurchPrice,\numsamp_{\optagent}) = \beta
\frac{\FishInv_{\optagent}}{\numsamp_{\optagent}} +
\PurchPrice\numsamp_{\optagent} = \p{1 +
  \FishInv_{\optagent}/\FishInvTil_{\optagent}}\sqrt{\beta
  \score_{(2)}}.
\end{align}
Under truthful reports from the sellers (i.e., $\FishInvTil_{\agentindex}
= \FishInv_{\agentindex}$), the principal's loss simplifies to $2
\sqrt{\beta s_{(2)}}$, so the principal achieves the same expected
loss as in Mechanism~\ref{mech:cost_only}. However, once agents may
strategically misreport $\FishInvTil_{\agentindex}$, the equivalence
no longer holds. It is important to understand how verification in
Step (6) of the protocol can restore truthful behavior.  In order to
build intuition, let us begin with an idealized case where the principal
can directly observe the true data quality \emph{ex post}.

\paragraph{Verification with Observed ex post Quality.}

Suppose that the principal can observe the true data quality
$\FishInv_{\optagent}$ of the winning agent after the data are
procured. In this case, Mechanism~\ref{mech:cost_and_quality} enforces
perfect verification of reported quality, and hence becomes fully
truthful.  Let us summarize this conclusion in a formal way:

\begin{proposition}
\label{proposition:DSIC_verfiable_quality}  
Under Mechanism~\ref{mech:cost_and_quality}, if the principal observes
the ground truth $\FishInv_{\optagent}$ after data procurement and
sets $\FishInvHat_{\optagent} = \FishInv_{\optagent}$, truthfully
reporting $(\price_\agentindex, \FishInvTil_\agentindex) =
(\cost_\agentindex, \FishInv_\agentindex)$ is a weakly dominant
strategy for every agent.
\end{proposition}
\noindent See~\Cref{sec:DSIC_verfiable_quality} for the proof.

This result shows that, under perfect \emph{ex post} verification,
Mechanism~\ref{mech:cost_and_quality} collapses to the truthful
scoring auction in Mechanism~\ref{mech:cost_only}. The principal can
effectively treat the data qualities $\FishInv_\agentindex$ as
truthfully reported and hence as known at the allocation stage.
Intuitively, when the principal can perfectly verify quality,
under-stating the inverse data quality $\FishInv_\agentindex$ is
punished by contract termination, while over-stating only increases
the amount of data the seller must supply at a fixed total
payment. Thus, sellers prefer to report their true data quality, and
we return to the truthful scoring auction of
Mechanism~\ref{mech:cost_only}.


\paragraph{Verification with Quality Estimated from Procured Data.}

In practice, however, the principal only sees a
finite sample of delivered data from the selected seller $\optagent$,
and can only form an estimate $\FishInvHat_\optagent$. Intuitively, one would hope that if this estimate is sufficiently
accurate, agents will still find it nearly optimal to report their
true quality. There are competing effects, however; agents can benefit from reporting higher quality through a better procurement score and a smaller required quantity, but they are also more likely to fail the verification step. As such, the agent's behavior is not immediately clear.

A natural way to reduce the risk of incorrectly rejecting an accurate quality report is to use a lenient verification statistic. For example, for a failure parameter $\alpha \in (0,1)$, one may use an approximate $(1-\alpha)$-level lower confidence bound $
Q_{\numsamp}(\alpha;\FishInv_\agentindex)$ satisfying
\begin{align}
\label{eq:LCB_bound}    
\PP{Q_n(\alpha; \FishInv_\agentindex) \leq \FishInv_\agentindex} &
\geq (1-\alpha) - \citol_\numobs,\qquad \citol_\numobs \to 0.
\end{align}
For large
$n$, a truthful report fails with probability at most about $\alpha$, and a small $\alpha$ corresponds to a more lenient test. 
In the Gaussian variance example of~\Cref{sec:simulation}, we discuss a concrete implementation of this test. 
The naive sample-variance test corresponds roughly to
$\alpha=1/2$, and may reject truthful reports too often. We give a conditional-utility calculation motivating this condition in~\Cref{sec:verf_estimated}.


\subsection{Approximate Incentives under Noisy Quality Verification}
\label{sec:incentive_analysis}

In this section, we formalize the noisy verification setting described in~\Cref{sec:mech_noisy} and study the reporting incentives induced by Mechanism~\ref{mech:cost_and_quality} when data quality can only be verified noisily from the procured samples. Unlike the exact-verification benchmark, truthful reporting need not be exactly optimal in finite samples. A seller may benefit from slightly exaggerating quality to improve their procurement score, or from reporting more conservatively to reduce the probability of failing verification. Moreover, because the seller incurs provision costs even when verification fails, noisy verification may also distort the reported cost.

We show, however, that these distortions vanish as the amount of procured data grows. Under mild regularity conditions, every seller whose type is bounded away from the upper boundary of the type space has a nearly truthful report that is approximately optimal against any realized profile of the other sellers' reports and yields nonnegative expected utility. This pointwise incentive guarantee in turn induces a near-truthful approximate Bayesian Nash equilibrium for any common-knowledge prior supported on uniformly interior types. Under this equilibrium, we show that the principal's regret decomposes into the usual efficiency loss of the known-quality second-score mechanism and an additional distortion from noisy verification that vanishes in the large-sample limit. Thus, as the amount of procured data grows, seller behavior approaches truthful reporting and the principal asymptotically recovers the performance of the corresponding known-quality mechanism.

\subsubsection{Asymptotically Truthful Reporting Incentives}
\label{subsec:incentives}

We start by formalizing a ``large-sample'' regime when the purchased sample size is
random and endogenous. We do this by considering a sequence of problem
instances indexed by an integer $k = 1,2,\dots$, with corresponding
accuracy weights $\beta_k > 0$. Recall from
equation~\eqref{eq:principal_loss} that $\beta$ governs how much the
principal values estimation accuracy relative to payment. 
We fix any
increasing sequence $\{\beta_k\}_{k=1}^\infty$ such that $\beta_k \to
\infty$, and refer to the environment with parameter $\beta = \beta_k$
as ``instance $k$''. 
Along any such sequence, and under
Assumption~\ref{assumption:compact}, the procured sample size chosen by
the mechanism is uniformly lower bounded by a deterministic sequence
(see~\Cref{lemma:sample_size_convergence}
in~\Cref{sec:sample_size_convergence}). Intuitively, as $k$ grows the
principal cares more about accuracy and the mechanism responds by
purchasing more samples, so that the procured sample size eventually
becomes large.

We impose the following boundedness condition on seller types and feasible reports. This assumption is standard in private-value Bayesian auction models~\citep{myerson1981optimal}.

\begin{assumption}
\label{assumption:compact}  
There exist constants $0 < \cunder < \cover < \infty$ and $0 <
\FishInvUnder < \FishInvOver < \infty$ such that, for each agent
$\agentindex \in [\numagent]$, their type $\type_\agentindex =
(\cost_\agentindex,\FishInv_\agentindex)$ takes values in $\TypeSet
\defn [\cunder, \cover] \times [\FishInvUnder, \FishInvOver]$.  Furthermore, agent
$\agentindex$'s reporting action $a_\agentindex = (\price_\agentindex,
\FishInvTil_\agentindex)$ is restricted to $\mathcal A = [\cunder,
  \cover] \times [\FishInvUnder, \FishInvOver]$.
\end{assumption}

Under the instance regime described above, we can now formalize the
informativeness of a verification test.  

\begin{assumption}
\label{assumption:test_tail_dens}  
There exists a family $\cb{\zeta_{n}}_{n=1}^\infty$ nonincreasing in both $n$ and its argument such that, conditionally on $(\FishInv, n)$, the tail probability
  \begin{align}
  \PP{\abs{\FishInvHat-\FishInv}\geq \delta\, \cond\, \FishInv, n}\le \zeta_{n}(\delta)  \qquad
    \mbox{for each $\delta > 0$,}
  \end{align}
and $\zeta_{n}(\delta)\to 0$ as $n\to\infty$ for every fixed $\delta>0$.
\end{assumption}

At a high level, Assumption~\ref{assumption:test_tail_dens} imposes only a mild requirement that the task-relevant quality admit a consistently estimable verification statistic. 
It requires that, as the number of procured samples grows, $\FishInvHat$ concentrates around the true quality $\FishInv$ uniformly over the feasible quality levels. 
For example, in the Gaussian variance setting with $\FishInvHat$ equal to the sample variance, the statistic follows an explicit scaled $\chi^2$ distribution, and standard concentration bounds yield an exponentially decaying tail envelope $\zeta_{n}(\delta)$.

The function $\zeta_{n}(\delta)$ provides an upper bound on the probability that the verification error exceeds $\delta$ when $n$ samples are procured. 
Consequently, a seller who exaggerates quality by at least $\delta$ passes verification with probability at most $\zeta_{n}(\delta)$, while a report that is conservative by $\delta$ fails with probability at most $\zeta_{n}(\delta)$. We provide further details of these verification guarantees in \Cref{sec:finite_dev}.

With this verification condition in place, we can now state our main incentive guarantee.
Let $\Util_\agentindex(a_\agentindex,
a_{-\agentindex}; \type_\agentindex)$ denote seller $\agentindex$'s
utility under the report profile $(a_\agentindex,
a_{-\agentindex})$ given type $\type_\agentindex=(\cost_\agentindex,\FishInv_\agentindex)$.
If seller $\agentindex$ is selected, then
\begin{equation}
\Util_\agentindex(a_\agentindex,
a_{-\agentindex}; \type_\agentindex)
=I(\FishInvHat\leq \FishInvTil)\cdot \bar{\price}_{\agentindex} \numsamp_{\agentindex} - \cost_{\agentindex}
\numsamp_{\agentindex},
\end{equation}
where $\bar{\price}_{\agentindex}$ and $\numsamp_{\agentindex}$ are the unit payment and procurement quantity induced by Mechanism~\ref{mech:cost_and_quality}.
Since both the seller's payment and provision cost scale with $\sqrt{\beta_k}$, we 
define the normalized utility
\begin{equation}
\util_{\agentindex,k}(a_\agentindex,
a_{-\agentindex}; \type_\agentindex):=\frac{\Util_\agentindex(a_\agentindex,
a_{-\agentindex}; \type_\agentindex)}{\sqrt{\beta_k}}.
\end{equation}
The following result shows that, against any fixed profile of the other sellers' reports, every interior seller has a nearly truthful report that is asymptotically optimal and yields nonnegative expected utility.

\begin{theorem}
\label{theorem:incentive}  
Under \Cref{assumption:compact} and \Cref{assumption:test_tail_dens},
there exist sequences $\tau_k, \lambda_k, \epsilon_k\to 0$ such that, for every seller $i$, every fixed type
$\type_\agentindex$ bounded away from the upper boundary of $\mathcal{T}$, and every fixed profile $a_{-\agentindex}$ of the other sellers' reports, for all sufficiently large $k$, the report
\begin{equation}
\aikDag := \p{\frac{\cost_\agentindex}{1-\lambda_k},\FishInv_\agentindex+\tau_k}
\label{eq:report}
\end{equation}
is feasible and satisfies 
\begin{enumerate}
\item[(a)] $\EE{\util_{\agentindex,k}(\aikDag,
a_{-\agentindex}; \type_\agentindex)}\ge0$,
  and
\item[(b)] $\sup_{a_\agentindex\in\mathcal A} 
\EE{\util_{\agentindex,k}(a_\agentindex,
a_{-\agentindex}; \type_\agentindex)}
- \EE{\util_{\agentindex,k}(\aikDag,
a_{-\agentindex}; \type_\agentindex)}\le \epsilon_k$,
\end{enumerate}
where the expectation is taken over the randomness of the verification statistic.
\end{theorem}
\noindent See~\Cref{subseq:incentive} for the proof, where we specify 
explicit constructions of $\tau_k,\lambda_k,\epsilon_k$ in
terms of the bounds in
\Cref{assumption:compact} and \Cref{assumption:test_tail_dens}.

\Cref{theorem:incentive} shows that, in the large-sample regime, every interior seller has a nearly truthful report that is approximately optimal. In particular, we show that (a) participating with the report $\aikDag$ yields nonnegative expected utility, and (b) $\aikDag$ achieves expected utility within $\epsilon_k$ of the best feasible report against any fixed profile of the other sellers' reports. Since $\tau_k, \lambda_k\to 0$, we also have $\aikDag\to(\cost_\agentindex,\FishInv_\agentindex)$, so both the cost and quality distortions vanish as the amount of procured data grows.

The form of $\aikDag$ reflects the uncertainty in finite-sample verification. Since $\FishInvHat_\agentindex$ is noisy, reporting the true quality $\FishInv_\agentindex$ exposes the seller to a nonzero probability of failing verification even when their report is accurate. The upward adjustment $\FishInv_\agentindex+\tau_k$ therefore provides a small conservative buffer. Assumption~\ref{assumption:test_tail_dens} allows $\tau_k$ to be chosen so that the corresponding failure probability is bounded by a vanishing quantity $\lambda_k$. 
Because the seller incurs the provision cost even when verification fails, the cost report is adjusted to ${\cost_\agentindex}/(1-\lambda_k)$ to compensate for this remaining verification risk. 
Thus, $\tau_k$ represents a vanishing conservative buffer in the reported quality, while $\lambda_k$ bounds the residual verification risk. The quantity $\epsilon_k$, in turn, measures the remaining utility gap between this near-truthful report and the seller's optimal strategic report.

To understand why $\aikDag$ is approximately optimal, consider the seller's utility in the corresponding known-quality benchmark.
We show that this benchmark upper bounds the utility attainable by any report up to a vanishing error. In particular, if a seller substantially exaggerates quality so that $\FishInvTil_\agentindex$ lies sufficiently below $\FishInv_\agentindex$,
Assumption~\ref{assumption:test_tail_dens} implies that the report passes verification only with small probability.
Otherwise, the claimed quality is close to or more conservative than the truth, and even if the report were guaranteed to pass verification, its utility could exceed the known-quality benchmark only by a vanishing amount. The proposed report $\aikDag$ also attains this benchmark up to a vanishing loss. Combining these two bounds yields the $\epsilon_k$-optimality guarantee.

\begin{remark}
The quantities $\tau_k$ and $\lambda_k$ in~\Cref{theorem:incentive} connect the incentive guarantee directly to the statistical accuracy of verification. At the effective sample size $N_k$, a quality deviation of $\tau_k$ can be detected with residual verification error bounded by $\lambda_k$. Thus, the quality buffer $\tau_k$ can be viewed as an effective finite-sample verification resolution at which these verification errors have already become small relative to the size of the deviation itself, while $\lambda_k$ measures the residual verification error at this resolution and determines the corresponding price markup. Together, they determine the incentive approximation error $\epsilon_k = \oo(\tau_k+\lambda_k)$.
For example, under a sub-Gaussian envelope $\zeta_\numsamp(\delta)\lesssim C_3 \exp\p{-C_4\numsamp \delta^2}$, the construction yields $\tau_k,\lambda_k,\epsilon_k = \oo\p{\sqrt{\log N_k/N_k} }$. We provide the explicit construction and further discussion in~\Cref{sec:finite_dev}.
\label{remark:finite_dev}
\end{remark}

\subsubsection{Approximate Equilibrium and Principal Regret}

The pointwise incentive guarantee in~\Cref{theorem:incentive} also has a natural equilibrium implication. To formalize this, suppose that seller types $(\BigType_1,...,\BigType_\numagent)$ are drawn from a common-knowledge prior $\prior$. Since the report $\aikDag$ in~\Cref{theorem:incentive} requires an upper buffer in both reported cost and inverse quality, we restrict attention to types uniformly bounded away from the corresponding upper boundaries. 
Specifically, let $\TypeSet_0\subset\TypeSet$ be compact with
\begin{equation*}
\sup_{(\cost,\FishInv)\in\TypeSet_0} c < \cover,\qquad
\sup_{(\cost,\FishInv)\in\TypeSet_0} \FishInv < \FishInvOver,
\end{equation*}
and suppose that $\prior$ is supported on $\TypeSet_0^\numagent$.
For instance $k$, a pure reporting strategy for seller $\agentindex$ is a measurable mapping $r_{\agentindex,k}:
\TypeSpace_0 \to \mathcal{A}^{\mathrm{out}}$, $\mathcal{A}^{\mathrm{out}}\defn \mathcal A \cup
\{\mathrm{out}\}$, where the action $\mathrm{out}$ corresponds to opting out and yields zero utility. We define below an approximate Bayesian Nash equilibrium.

\begin{definition}[$\epsilon$-Bayesian Nash equilibrium]
A strategy profile $r_k \defn (r_{1k}, \ldots, r_{\numagent k})$
is an \emph{$\epsilon$-Bayesian Nash equilibrium} for instance $k$ if, for every seller
$\agentindex \in [\numagent]$ and $\prior_i$-almost every type $\type_\agentindex \in \TypeSpace_0$,
\begin{equation}
\EE{\util_{\agentindex,k}(r_{ik}(\type_i),
r_{-i,k}(\BigType_{-i}); \type_\agentindex)\cond \BigType_\agentindex=t_\agentindex}
\ge 
\sup_{a_\agentindex\in\mathcal A^{\mathrm{out}}} 
\EE{\util_{\agentindex,k}(a_\agentindex,
r_{-i,k}(\BigType_{-i}); \type_\agentindex)\cond \BigType_\agentindex=t_\agentindex}
- \epsilon
\end{equation}
where $\prior_i$ denotes the marginal distribution of $\BigType_\agentindex$, and the expectation is taken over the conditional distribution of the other sellers' types $\BigType_{-\agentindex}$ given $\BigType_\agentindex=t_\agentindex$ as well as the randomness of the verification statistic.
\end{definition}

We now consider the near-truthful strategy induced by~\Cref{theorem:incentive}. For each seller $\agentindex$, 
define the strategy 
\begin{equation}
\rikDag(\cost_\agentindex,\FishInv_\agentindex)\defn
\p{\frac{\cost_\agentindex}{1-\lambda_k},\FishInv_\agentindex+\tau_k},\qquad
(\cost_\agentindex,\FishInv_\agentindex)\in\TypeSet_0.
\end{equation}
Since $\TypeSet_0$ is uniformly bounded away from the upper boundary and $\tau_k,\lambda_k\to 0$, $\rikDag(\type_\agentindex)$ is feasible for every $\type_\agentindex\in \TypeSet_0$ for all sufficiently large $k$.
Let $\rkDag\defn (r_{1k}^{\dagger},...,r_{Mk}^{\dagger})$ be the associated strategy profile.
The pointwise guarantee in~\Cref{theorem:incentive} immediately yields the following equilibrium consequence.

\begin{coro}
\label{coro:BNE}  
Under~\Cref{assumption:compact} and~\Cref{assumption:test_tail_dens},
for all sufficiently large $k$, $\rkDag$ is an $\epsilon_k$-approximate Bayesian Nash equilibrium, where $\epsilon_k\to 0$ is as defined in~\Cref{theorem:incentive}. Moreover, every seller obtains nonnegative interim expected utility under $\rkDag$.
\end{coro}
\noindent See~\Cref{subseq:BNE} for the proof.

Corollary~\ref{coro:BNE} translates the pointwise incentive guarantee of Theorem~\ref{theorem:incentive} into equilibrium behavior. Under $\rkDag$, every seller adopts the near-truthful report identified in Theorem~\ref{theorem:incentive}, and hence both the reported cost and quality converge to their true values as $k\to\infty$. Thus, as the amount of procured data grows, the strategic behavior induced by noisy verification approaches that of the known-quality mechanism.

We next examine what this near-truthful equilibrium implies for the principal. 
Formally, let $\score^\truth_\agentindex \defn \cost_\agentindex \FishInv_\agentindex$ denote the truthful score of agent $\agentindex$, with order statistics $\score^\truth_{(1)}\le \score^\truth_{(2)}$, 
let $\loss_k^\FB\defn 2\sqrt{\beta \;\score^\truth_{(1)}}$ denote the first-best loss, and let $\loss_k$ denote the principal's realized loss\footnote{We take the principal to retain the delivered data when verification fails. If instead the principal discards the failed data and uses a standard fallback procurement option, the same asymptotic conclusion holds in expectation, since failure under $\rkDag$ occurs with probability at most $\lambda_k$.}  under~\Cref{mech:cost_and_quality}.
The following proposition characterizes the principal's regret under the near-truthful equilibrium $\rkDag$.

\begin{proposition}
\label{proposition:principal_loss}
Under~\Cref{assumption:compact} and~\Cref{assumption:test_tail_dens} and with the strategy profile $\rkDag$,
\begin{equation}
\frac{\loss_k-\loss_k^\FB}{\loss_k^\FB} \le \underbrace{\sqrt{\frac{\score^\truth_{(2)}}{\score^\truth_{(1)}}}-1}_{\text{second-score competition gap}}+ \underbrace{\oo\p{\tau_k+\lambda_k}}_{\text{additional distortion from noisy verification}}.
\end{equation}
\end{proposition}
\noindent See~\Cref{subsec:principal_loss} for the proof.

\Cref{proposition:principal_loss} decomposes the principal's regret into two sources. The first term is the usual competition gap of the second-score mechanism and is present even when seller quality is known. The second term captures the additional loss induced by noisy verification and strategic reporting; unlike the competition gap, this distortion vanishes as the procured sample size grows. In particular, under the sub-Gaussian verification setting of~\Cref{remark:finite_dev}, the additional regret is
$\oo\p{\sqrt{\log N_k/N_k} }$. Thus, asymptotically, noisy verification introduces no additional first-order loss relative to the corresponding known-quality mechanism.

\section{Numerical results}
\label{sec:cifar}

We explore the behavior of our mechanism for a semi-synthetic form of classifier-evaluation setting based on the CIFAR-10H dataset~\citep{krizhevsky2009learning,peterson2019human}.
In~\Cref{sec:simulation}, we also give a simple Gaussian variance illustration of the mechanism, where the verification statistic can be written explicitly and the qualitative incentive patterns highlighted by our theory can be seen in a more transparent setting.

Consider the classifier-evaluation problem: a buyer has a classifier $h$ and wants to estimate its accuracy relative to a population of human labels. 
For an image $X$, let $\PP{Y=y\cond X}$ denote the distribution of human responses. The buyer's goal is to estimate the scalar
$\theta\defn \EE[X]{p_h(X)}$ where $p_h(X)\defn \PP{Y=h(X)\cond X}$,
corresponding to the average probability that the classifier $h$ agrees with a (randomly sampled) human label.

Each seller $\agentindex$ corresponds to a labeling protocol.
Sellers differ in the effective number of independent human labels they provide per image. Although the raw number of labels can be counted, the effective number of independent labels is not directly verifiable, as a provider may reuse labels or rely on highly correlated raters.
We summarize this after accounting for redundancy or dependence across labels by an effective label multiplicity $m_i$. For an image $X_{\agentindex,t}$, let $A_{\agentindex,t}\in[0,1]$ denote the image-level agreement score with the classifier. We assume that
\begin{equation}
\EE{A_{\agentindex,t}\cond X_{\agentindex,t}} = p_h(X_{\agentindex,t}),\qquad \mbox{and} \quad
\Var{A_{\agentindex,t}\cond X_{\agentindex,t}} = \cb{p_h(X_{\agentindex,t})(1-p_h(X_{\agentindex,t}))}/{m_i}.
\end{equation}
The buyer then estimates the classifier’s soft human accuracy with $\htheta_i\defn \sum_{t=1}^{n_i} A_{\agentindex,t} / n_i$.
Note that $\htheta_i$ is an unbiased estimator of $\theta$, but with different variance $\text{Var}\,(\htheta_i) = \FishInv_i/\numsamp_i$, where $\FishInv_i\equiv \Var{p_h(X_{\agentindex,t})} + \EE{\Var{A_{\agentindex,t}\cond X_{\agentindex,t}}}$.

Since the inverse quality $\FishInv_i$ is task-specific, sellers may not be able to report it directly. We therefore ask seller $\agentindex$ to 
report the effective number of independent labels $\tilde m_i$. Using a public reference set $\mathcal D_{\rm ref}$, the buyer estimates $p_h$ and calibrates a map $m\mapsto V_{\mathcal D_{\rm ref}}(m)$. The reported multiplicity $\tilde m_i$ is then converted into the reported inverse quality $\FishInvTil_\agentindex :=V_{\mathcal D_{\rm ref}}(\tilde m_i)$, which is used in Mechanism~\ref{mech:cost_and_quality}.
After procurement, the buyer uses the delivered data to estimate a delivered multiplicity $\hat m_i$, and verifies the report by comparing $\FishInvHat_\agentindex\defn V_{\mathcal D_{\rm ref}}(\hat m_i)$ with $\FishInvTil_\agentindex$.

We use CIFAR-10H to construct both the distribution of images and the conditional distribution of human responses, which we treat as ground truth.
For a fixed classifier $h$, the principal observes only a small reserved reference subset, on which she estimates $p_h(x)$ by the empirical fraction of CIFAR-10H labels agreeing with $h(x)$, and uses these estimates to calibrate the map. 
We consider two verification rules for the delivered multiplicity: (i) a ``Point Estimate'' rule, which uses a method-of-moments estimator $\hat m_i$ computed from the delivered data; and (ii) a ``95\% UCB'' rule, which replaces $\hat m_i$ by its 95\% upper confidence bound before applying the verification step,\footnote{We use a UCB rather than an LCB because $m_\agentindex$ is the quality parameter, while the mechanism is written in terms of inverse quality $\FishInv_\agentindex$, which decreases with $m_\agentindex$. } and compare them to the benchmark of no verification. Further simulation details can be found in~\Cref{sec:details_cifar}.

\begin{figure}[t]
\centering
\includegraphics[width=0.9\linewidth]{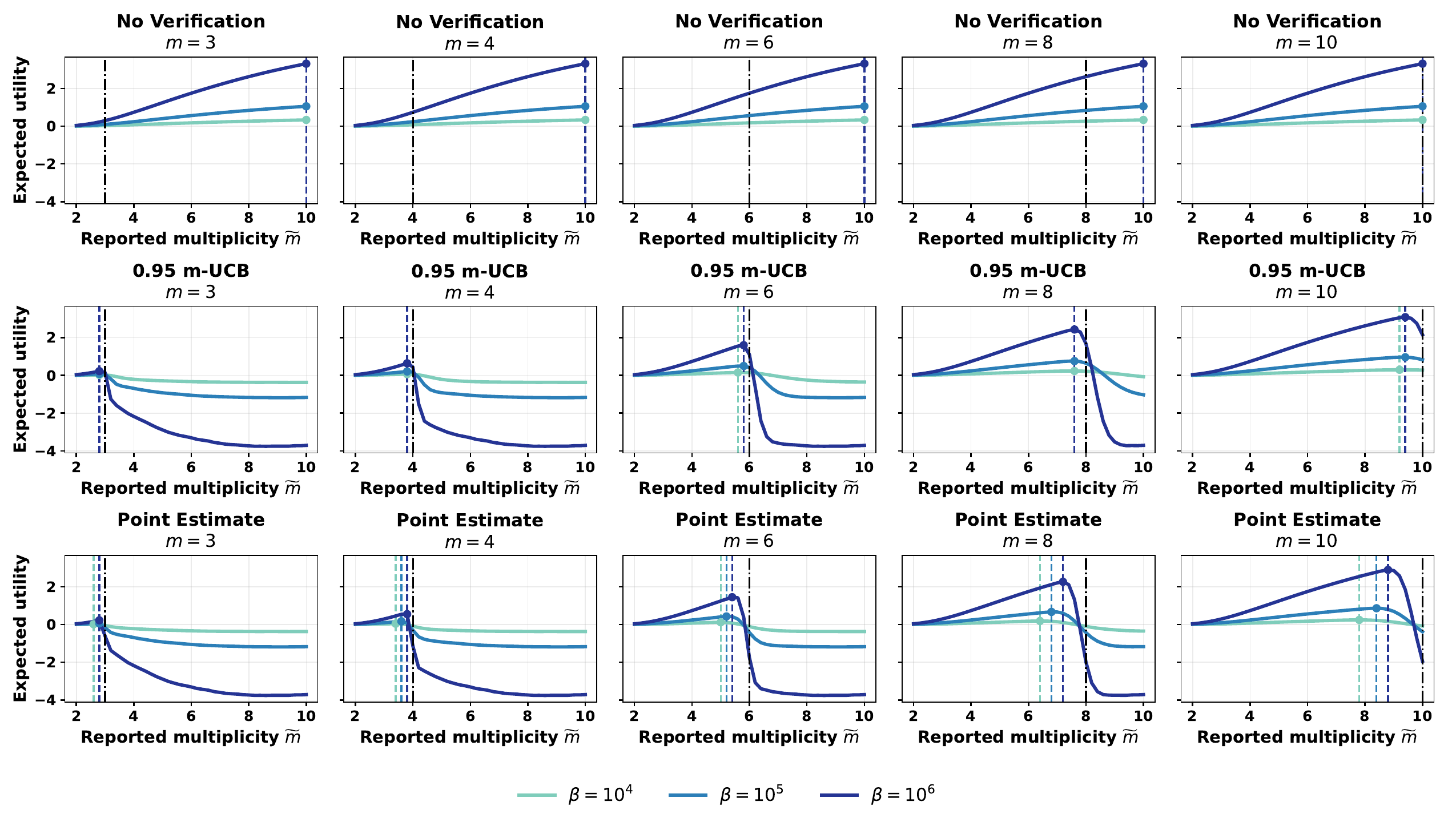}
\caption{Interim expected utility of the focal agent as a function of
  the reported multiplicity, for different true multiplicities (columns),
  verification rules (rows), and loss parameters (colors). The black
  dot–dashed line marks the true multiplicity; colored dashed lines and
  dots indicate, for each $\beta$, the report that maximizes interim
  expected utility.  }
\label{fig:cifar_utility}
\end{figure}

We examine the agent’s interim expected utility and the
corresponding optimal reporting behavior when the other agents report
truthfully. To isolate the quality-reporting incentives, we fix a focal agent, hold its reported cost at the true cost, and evaluate its interim expected utility as a function of the reported multiplicity $\tilde m_\agentindex$. \Cref{fig:cifar_utility} plots the resulting utility curves for several true multiplicities, verification rules, and values of the tradeoff parameter. See~\Cref{sec:details_cifar} for additional details.

Without verification, the focal agent has a clear incentive to over-report quality as much as possible. However, once verification is introduced, the utility-maximizing report moves much closer to the truth across all settings.
The less strict 95\% UCB rule tends to induce very mild under-reporting, whereas the stricter Point Estimate rule leads to much more conservative under-reporting.
As $\beta$ increases, the utility-maximizing quality report moves closer to the true multiplicity across the verification rules. This complements our theoretical result in~\Cref{sec:incentive_analysis}, which guarantees the existence of asymptotically optimal near-truthful reports as the procured sample size grows. In our numerical example, the exact utility-maximizing quality reports themselves also move toward the truth. 
Moreover, the more lenient 95\% $m$-UCB rule approaches truthful reporting considerably faster than the Point Estimate rule.
This suggests that, in our example, a lenient test is more effective for truth elicitation at finite $\beta$.

\section{Discussion}
\label{SecDiscussion}


We take a step toward a theory of data procurement as
information acquisition under limited prior knowledge of providers'
costs and data quality.  Our mechanism combines a selection rule from
second-score procurement auctions with an enforcement device in the
spirit of scoring-rule-based forecast evaluation.  By leveraging
competition and a noisy ex post signal about the information
delivered, it induces approximately optimal near-truthful reporting and individually rational participation in large-sample regimes---something that is hard to obtain with
purely auction-based or purely scoring-rule-based mechanisms.

Our analysis has focused on a scalar information setting, where each
agent's data quality is captured by a one-dimensional information
metric, such as how noisy their data are.  In this case, for any fixed
pricing scheme, the purchasing rule that minimizes the principal's
loss is a winner-take-all one, where it is optimal to buy all samples
from a single agent even though the principal could, in principle, buy
from multiple agents. This property inherently enables us to design a
simple second-score mechanism that always ensures an optimal
procurement quantity.  In many applications, however, the optimal
design would involve purchasing data from multiple agents, for
example, when the parameter of interest is vector-valued and different
agents are informative about different components or regions of the
covariate space, or when the principal wants to guard against
source-specific biases by combining data from multiple agents.
Extending our framework to such settings, where the optimal allocation
may buy positive quantities from multiple agents rather than always
from a single source, would require mechanisms that jointly determine
prices and quantities across all data providers, and we leave this as
an interesting direction for future work.

In richer settings, data quality may not admit a pre-specified parametric form, and sellers may not know how informative their data are for the buyer’s estimation objective. Instead of asking sellers to report an information parameter, one could ask them to provide richer evidence of data quality, such as a pilot dataset or task-relevant summary statistics. The buyer could then use the delivered data to test whether this ex ante evidence is representative of the data ultimately supplied. Developing such mechanisms would extend the verification idea beyond information measures and toward more flexible data-procurement settings. 

\myprocbibliography

\myprocappendix

\section{Extensions to Non-Root-$n$ Losses}
\label{sec:sub_loss}

In this section, we discuss how our information procurement mechanisms
can be extended to the cases where the estimation cannot in general
achieve a parametric rate. Specifically, we consider a setting where
the statistician (i.e., the principal) can estimate the parameter of
interest up to mean-squared accuracy of
\begin{align}
\label{eq:mse_form_sub}
\operatorname{MSE} & \asymp \p{\frac{1}{ \numobs_\agentindex
    \Fish_\agentindex}}^\rho \equiv
\p{\frac{\InvFish_\agentindex}{\numobs_\agentindex}}^\rho
\end{align}
for some problem-dependent exponent $\rho\in(0,1)$, in which case her
loss is given by the weighted sum
\begin{align}
\label{eq:principal_loss_sub}
  \Loss(\agentindex, \price_{\agentindex},\numsamp_{\agentindex}) &
  \defn \beta \cdot
  \p{\frac{\FishInv_{\agentindex}}{\numsamp_{\agentindex}}}^\rho +
  \price_{\agentindex} \numsamp_{\agentindex}.
\end{align}
In this case, $\InvFish_\agentindex$ is a generalized inverse
information parameter that captures the per-sample inverse
information.  Such errors are prevalent, for example, in nonparametric
regression and density estimation, where $V_i$ reflects agent-specific
data quality such as noise level and covariate spread, while $\rho$
depends on problem-specific features such as smoothness and
dimensionality.

Mechanism~\ref{mech:cost_only_sub} below extends
Mechanism~\ref{mech:cost_only} to sample purchasing in these
generalized loss settings. Since $\InvFish_\agentindex$ continues to
capture per-sample inverse information, minimizing the buyer’s loss is
still equivalent to minimizing $\price_i\InvFish_\agentindex$.
Therefore, the scoring rule, the selection of the winner, and the unit
pricing rule need not be altered and remain the same as in
Mechanism~\ref{mech:cost_only}.  The only change is in Step~5, which
now sets the purchased quantity to the optimal sample size
corresponding to the unit payment $\price_\optagent$ under the
generalized loss, so that the principal still buys an optimal amount
of data at the actual price she pays.

\mygraybox{
  \begin{mech}
Second-Price-per-Information Mechanism.
\begin{enumerate}
\item \textbf{Bidding}: Each agent $\agentindex \in [\numagent]$
  decides whether to enter the mechanism, leading the \emph{opt-in
  set} $\optinset \subseteq [\numagent]$ of participating agents. Each
  opt-in agent $\agentindex \in\optinset$ submits a per-sample price
  bid $\price_\agentindex$.
\item \textbf{Scoring}: The principal computes the scores
  $\score_\agentindex \defn \price_\agentindex \FishInv_\agentindex$
  for each $\agentindex \in \optinset$.
  \item \textbf{Selection}: The principal chooses the agent $\optagent
    \defn \arg \min \limits_{\agentindex \in \optinset}
    \score_\agentindex$ with the lowest score.
\item \textbf{Unit Payment}: Letting $s_{(2)} \defn \min
  \limits_{\agentindex\in \optinset \backslash \{\optindex \}}
  \score_\agentindex$ denote the smallest  score among the losing
  agents, the winner $\optagent$ is paid a per-sample price
\begin{subequations}
\begin{align}
\PurchPrice \defn \frac{s_{(2)}}{\FishInv_{\optagent}} \; = \; s_{(2)} \Fish_{\optagent}
\end{align}
corresponding to the second-best score scaled by the winner's Fisher information.
\item \textbf{Quantity}: The principal purchases a total of
\begin{align}
\numsamp_{\optagent} = \p{\frac{\beta\rho}{
    s_{(2)}}}^{\frac{1}{\rho+1}} \FishInv_{\optagent} \qquad \text{samples from seller $\optagent$.}
\end{align}
\end{subequations}
\end{enumerate}
\label{mech:cost_only_sub}
\end{mech}
}

To provide intuition, let's still focus on the benchmark setting where the buyer observes each seller’s true quality $\FishInv_\agentindex$ ex ante.
It is not hard to verify that,
under Mechanism~\ref{mech:cost_only_sub}, the principal achieves a total loss of
\begin{align*}
\Loss(\optagent, \PurchPrice, \numsamp_{\optagent}) = (1+\rho)\rho^{-\rho/(\rho+1)}\beta^{1/(\rho+1)} \p{\score_{(2)}}^{\rho/(\rho+1)}.
\end{align*}
This is exactly the optimal loss that would be achieved if the principal could purchase optimally from the second-best agent with the true cost. It gives rise to a gap from the first-best solution
\begin{align*}
(1+\rho)\rho^{-\rho/(\rho+1)}\beta^{1/(\rho+1)} \cb{\p{\score_{(2)}}^{\rho/(\rho+1)} - \p{\score_{(1)}}^{\rho/(\rho+1)}},
\end{align*}
which is entirely driven by the optimal-loss gap between the first-best and second-best agents. Since Mechanism~\ref{mech:cost_only_sub} does not use the winner's own report to determine either the unit price or the purchased quantity, it retains the same truthful and individually rational properties as Mechanism~\ref{mech:cost_only} in the benchmark setting, in line with standard second-score mechanisms.

For noisy verification, we expect the main conclusion of Section~\ref{sec:incentive_analysis} to continue to hold under the general scaling. In particular, under the same verification regularity condition, there should exist sequences $\tau_k,\lambda_k,\epsilon_k\to 0$ such that every interior seller has a near-truthful report $\p{\cost_\agentindex/\p{1-\lambda_k},\FishInv_\agentindex+\tau_k}$ that is $\epsilon_k$-optimal against any fixed profile of the other sellers' reports and yields nonnegative expected utility. The argument is essentially unchanged, except that seller utility is now naturally normalized by $\beta^{1/(\rho+1)}$, and the lower bound on the procured sample size scales as $N_k\asymp \beta^{1/(\rho+1)}$. 
Consequently, the finite-sample rates of $\tau_k,\lambda_k,\epsilon_k$ would be determined by the verification envelope evaluated at this modified sample-size scale.
As $\beta_k\to\infty$, both reported cost and quality again converge to the truth, with an analogous near-truthful approximate equilibrium and vanishing additional principal-side distortion.


\section{Verification with Quality Estimated from Procured Data}
\label{sec:verf_estimated}

In~\Cref{sec:mech_noisy}, we established that
Mechanism~\ref{mech:cost_and_quality} is fully truthful if the
principal is given exact knowledge of winner’s data quality
$\FishInv_\optagent$. In practice, however, the principal only sees a
finite sample of delivered data from the selected seller $\optagent$,
and can only form an estimate $\FishInvHat_\optagent$. In this section, we aim to build up intuition for impact of imperfect verification.

Intuitively, one would hope that if this estimate is sufficiently
accurate, agents will still find it nearly optimal to report their
true quality. There are competing effects, however; agents can benefit from reporting higher quality through a better procurement score and a smaller required quantity, but they are also more likely to fail the verification step. As such, the agent's behavior is not immediately clear.


To gain insight, we begin a simple calculation of conditional
individual-rationality.  Fix a seller $\agentindex$, and let
$\score_{-\agentindex,(1)}$ denote the smallest score among the other
sellers who opt in.  Conditionally on $\score_{-\agentindex,(1)}$,
suppose that agent $\agentindex$ reports the pair
$(\price_\agentindex, \FishInvTil_{\agentindex})$ and wins the auction
with score $\score_\agentindex(\price_\agentindex,
\FishInvTil_{\agentindex}) = \price_\agentindex
\FishInvTil_{\agentindex} < \score_{-\agentindex,(1)}$.  The
(conditional) winning utility for agent $\agentindex$ is therefore
\begin{subequations}
\begin{align}
\Util_\agentindex(\price_\agentindex, \FishInvTil_\agentindex) =
\PP{\FishInv_{\agentindex} \leq \FishInvTil_{\agentindex}} \cdot
\sqrt{\beta \score_{-\agentindex,(1)}} - \frac{\sqrt{\beta}
  \cost_\agentindex \FishInvTil_\agentindex}{\sqrt{
    \score_{-\agentindex,(1)}}}.
\label{eq:winning_utility}
\end{align}
Under truthful reporting, $(\price_\agentindex,
\FishInvTil_{\agentindex}) = (\cost_\agentindex,
\FishInv_{\agentindex})$, this expected utility is non-negative only
if
\begin{align}
\label{eq:conditional_opt_in_cond}  
\PP{\FishInvHat_{\agentindex} \leq \FishInv_{\agentindex}} \ge \frac{
  \score_\agentindex(\cost_\agentindex, \FishInv_\agentindex)
}{\score_{-\agentindex,(1)} }.
\end{align}
\end{subequations}
where $\score_\agentindex(\cost_\agentindex, \FishInv_\agentindex) =
\cost_\agentindex \FishInv_\agentindex$ is agent $\agentindex$'s true
score. In words, conditional on the rivals' scores, a truthful agent
finds participation individually rational only if the probability of failing verification is sufficiently low relative to the separation between their bid and that of the second best. 


We now examine this in more detail with a naive verification strategy.
Consider a Gaussian location model, for which the (inverse) data
quality $\FishInv_\agentindex \equiv \sigsq_\agentindex$ corresponds
to the variance of the observations.  A natural estimate of the
variance $\sighatsq_\agentindex$ is the sample variance based on the
$\numsamp_{\optagent}$ procured samples. By classical large-sample
theory~\citep{vaart1998asymptotic}, the sample variance
$\sighatsq_\agentindex$ falls below the true value with probability
close to one half for large $\numsamp_{\optagent}$---that is, we have
\begin{align}
  \label{eq:sample_mean_failure}
\PP{\sighatsq_{\agentindex} \leq \sigsq_{\agentindex}} - \frac{1}{2} &
= \oo\p{\frac{1}{\sqrt{\numsamp_{\optagent}}}}.
\end{align}
Consequently, when the ratio between the first-best and second-best
scores is larger than $1/2$, a truthful best-score seller would
expect negative utility from winning.  Thus, such an agent should find
participation unattractive whenever their score is not sufficiently
smaller than the smallest competing score.


\section{Finite-Sample Incentive Bounds}
\label{sec:finite_dev}

In~\Cref{sec:incentive_analysis}, we establish the existence of vanishing sequences $\tau_k, \lambda_k, \epsilon_k\to 0$ that control the reporting and incentive distortions induced by noisy verification. In this section, we make these quantities explicit and illustrate how their finite-sample behavior depends on the concentration of the verification statistic and the amount of data procured by the mechanism.

Recall that $N_k$ denotes the deterministic lower bound on the procured sample size in instance $k$, with $N_k\to\infty$ (see~\Cref{lemma:sample_size_convergence} for details). Under~\Cref{assumption:test_tail_dens}, one valid construction is
\begin{subequations}
\begin{equation}
\tau_k\defn\inf\cb{\delta>0:\zeta_{N_k}(\delta)\le \delta } +\frac{1}{N_k},\qquad
\lambda_k\defn\zeta_{N_k}(\tau_k)
\label{eq:tau_lambda_construction}
\end{equation}
and the proof of~\Cref{theorem:incentive} gives
\begin{equation}
\epsilon_k\defn (C_1+C_2)(\tau_k+\lambda_k)\to 0.
\end{equation}
\end{subequations}
for some constants $C_1,C_2<\infty$ depending only on the bounds in~\Cref{assumption:compact}. 

The construction in~\Cref{eq:tau_lambda_construction} links the incentive guarantee directly to the statistical accuracy of verification. With at least $N_k$ samples,~\Cref{assumption:test_tail_dens} implies that a quality exaggeration of size $\delta$ passes verification with probability at most $\zeta_{N_k}(\delta)$, while a conservative report $\FishInv+\delta$ fails with probability at most the same quantity. The value $\tau_k$ therefore identifies a quality scale at which these verification errors have already become small relative to the size of the deviation itself. In this sense, $\tau_k$ can be viewed as an effective finite-sample resolution of the verification problem, while $\lambda_k=\zeta_{N_k}(\tau_k)$ bounds the residual verification error at that resolution.

The bound~\Cref{eq:tau_lambda_construction} then translates these statistical quantities into the seller's incentive loss. Sharper concentration of the verification statistic permits a smaller resolution $\tau_k$, a smaller residual verification error $\lambda_k$, and consequently a smaller approximation error $\epsilon_k$. We next make these rates explicit under a sub-Gaussian verification envelope.

\subsection{Finite-Sample Incentive Bounds under Sub-Gaussian Verification}

Suppose that the verification statistic satisfies~\Cref{assumption:test_tail_dens} with an envelope of the form
\begin{equation}
\zeta_\numsamp(\delta)\lesssim C_3 \exp\p{-C_4\numsamp \delta^2},
\end{equation}
for $\delta$ in a neighborhood of zero and constants $C_3,C_4>0$. Such an envelope corresponds to the usual $n^{-1/2}$ statistical concentration scale. To upper bound the construction of $\tau_k$ in~\Cref{eq:tau_lambda_construction}, take
\begin{equation*}
\delta_\numsamp= C_5 \sqrt{\frac{\log\numsamp}{\numsamp}}
\end{equation*}
for a sufficiently large constant $C_5$ satisfying $C_4C_5^2>1/2$. Then 
\begin{equation*}
\zeta_\numsamp(\delta_\numsamp) \le C_3 n^{-C_4C_5^2} \le C_5 \sqrt{\frac{\log\numsamp}{\numsamp}}=\delta_\numsamp.
\end{equation*}
Thus, $\delta_n$ is feasible in the infimum defining $\tau_k$, and therefore
\begin{equation*}
\tau_k \le C_5 \sqrt{\frac{\log N_k}{N_k}} + \frac{1}{N_k} = \oo\p{\sqrt{\frac{\log N_k}{N_k}} }.
\end{equation*}
Moreover, by construction $\lambda_k=\zeta_{N_k}(\tau_k)\le \tau_k$, and thus
\begin{equation*}
\lambda_k,\epsilon_k = \oo\p{\sqrt{\frac{\log N_k}{N_k}} }.
\end{equation*}

The Gaussian variance model provides a direct example. In the Gaussian location model, the inverse Fisher information is $\FishInv\equiv \sigma^2$, and the verification statistic based on the sample variance concentrates around $\FishInv$ at the usual $\numsamp^{-1/2}$ scale. Indeed, since the sample variance has a scaled $\chi^2$ distribution, \Cref{assumption:test_tail_dens} is satisfied with a sub-Gaussian envelope of the form above uniformly over the compact variance space. Thus, the quality buffer and normalized incentive gap shrink at the near-parametric rate $\sqrt{\log N_k/N_k}$.

The same framework also applies to the lenient LCB rule. In the Gaussian variance model, the LCB differs from the sample variance by an $\oop(\numsamp^{-1/2})$ conservative adjustment. Hence it remains a consistent verification statistic and satisfies~\Cref{assumption:test_tail_dens}. In particular, the adjustment occurs at the same first-order $\numsamp^{-1/2}$ statistical scale as the sampling error of the variance estimator. This explains why the LCB retains vanishing reporting and incentive distortions while offering a more lenient finite-sample verification rule.


\section{Simulation Details and Additional Numerical Results}
\label{sec:sim}

\subsection{Additional simulation details for the CIFAR-10H example}
\label{sec:details_cifar}

In the simulations, we fix $\numagent=10$ agents, draw $$c_i\sim \text{Uniform}(0.001,0.01),\qquad m_i\sim \text{Uniform}(2,10),$$ and generate image-level agreement variables $$A_{i,t}\cond X_{i,t}=x \sim {\rm Beta}((m_i-1)p_h(x), (m_i-1)(1-p_h(x)) ).$$
We consider a semi-synthetic classifier $h$ by taking the modal CIFAR-10H label and perturbing it independently with probability $0.2$, in which case the prediction is replaced by a uniformly sampled non-modal class.
To implement the mechanism, we reserve a random disjoint subset of size 200 as the public reference set.

For examining the optimal reporting behavior, we fix a focal agent $\agentindex$
with $\cost_\agentindex=0.0015$ and
$m_\agentindex\in \cb{3,4,6,8,10}$, and evaluate the expected utility as a function of $\tilde m_\agentindex\in [2,10]$ for three values of the tradeoff parameter $\beta\in\cb{10^4, 10^5, 10^6}$. We approximate the acceptance probability of each verification rule using the asymptotic normal distribution of the sample-variance statistic, with variance determined by the corresponding fourth moment; the outer expectation over competing sellers is approximated using 5,000 Monte Carlo draws.

\subsection{Participation Incentives under Truthful Cost Reporting}
\label{sec:sim_participation}

\begin{figure}[t]
\centering
\includegraphics[width=0.9\linewidth]{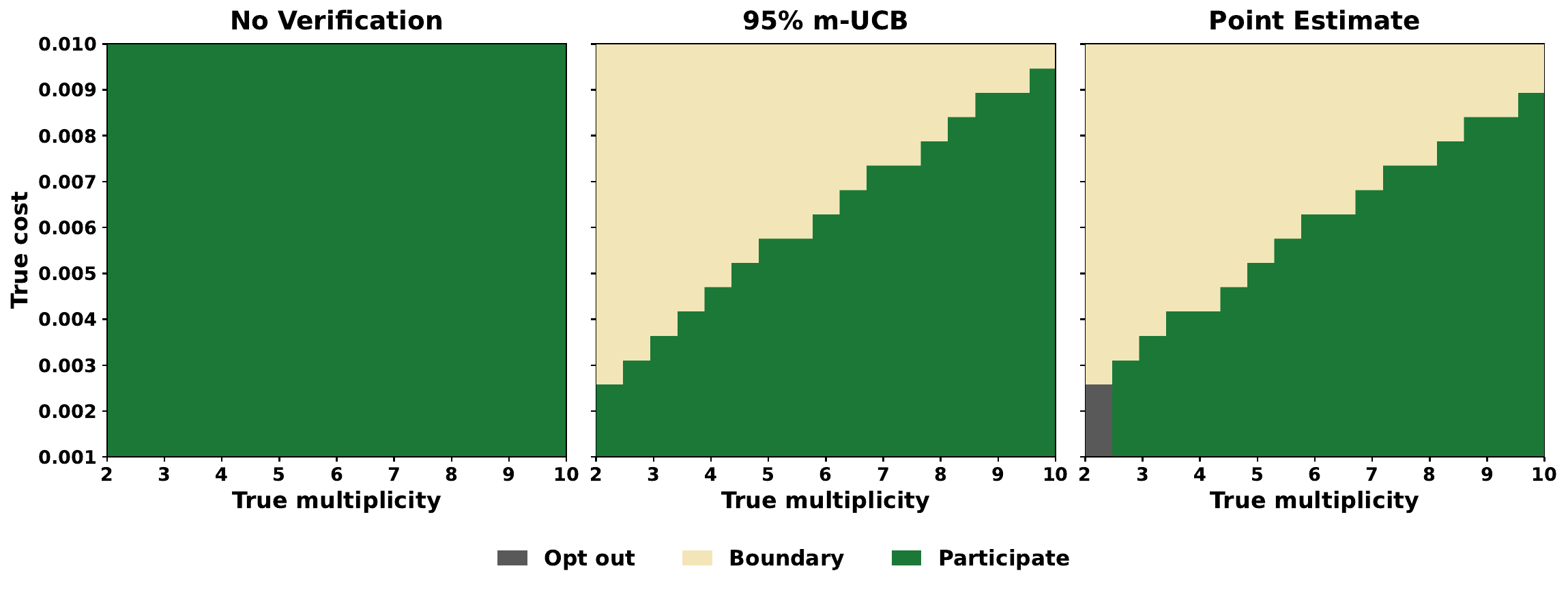}
\caption{Participation incentive as a
  function of type $(\cost_\agentindex, m_\agentindex)$ for
  $\beta=10^6$, with no verification rule (left), the 95\% UCB rule (middle), and the Point Estimate rule (right). Green indicates types that prefer participating, grey indicates types that prefer opting out, and beige indicates boundary types with near-zero interim winning utility.}
\label{fig:cifar_multiplicity_participation}
\end{figure}

We additionally examine finite-sample participation incentives in~\Cref{sec:cifar} when the focal seller's cost report is fixed at its true cost. We fix $\beta=10^6$ and, for each test, evaluate a grid of types $(\cost_\agentindex,m_\agentindex)$. For each grid point, we compute the focal agent’s interim expected winning utility at their optimal multiplicity report (given truthful reporting by the other agents), and compare this to the outside option of not participating, which yields zero. 
The participation incentives can be found in~\Cref{fig:cifar_multiplicity_participation}.
This participation map
illustrates that, when cost reporting is held truthful, lenient verification substantially enlarges the finite-sample participation region. This also explains the small cost adjustment in~\Cref{theorem:incentive}, which compensates sellers for residual verification risk.

\subsection{Numerical illustration with a Gaussian location model}
\label{sec:simulation}

In this section, we provide a concrete Gaussian model example to
illustrate how the verification test can be implemented in practice
and what reporting incentives it induces. Throughout, we consider
$\numagent = 10$ symmetric agents with types $\type_\agentindex =
(\cost_\agentindex, \FishInv_\agentindex)$ drawn i.i.d. from a common
prior with compact support, where $\cost_\agentindex \sim
\text{Uniform}(0.1,0.2)$ and $\FishInv_\agentindex \sim
\text{Uniform}(10,20)$. Conditional on type, agent $\agentindex$’s data
$X_{i,t}$ are i.i.d. from a normal distribution
$N(0,\sigsq_\agentindex)$. 
In this Gaussian example, $\FishInv_\agentindex$ coincides with the variance
$\sigsq_\agentindex$, and for notational convenience we will write $\sigsq_\agentindex$ in what follows.

\subsection{Verification Tests}

To construct the test, we need a statistic $\sighat_{\optagent}^2=
Q_{n_{\optagent}}(\alpha;\sigma_{\optagent}^2)$ such that Assumption~\ref{assumption:test_tail_dens}
is satisfied.  It is possible to construct such a statistic from the
sample variance. In particular, we define
\begin{equation}
S_n^2 \defn \frac{1}{n}\sum_{t=1}^n (X_t-\bar X)^2, \quad \mbox{and}
\quad \hat \mu_{4,n} \defn \frac{1}{n}\sum_{t=1}^n (X_{t}-\bar X)^4,
\end{equation}
where $\bar X$ is the sample mean of the procured data.  Under a
finite sixth moment, a one-sided normal approximation for the sample
variance yields the lower confidence bound (LCB)
\begin{equation}
Q_n(\alpha;\sigma^2) = S_n^2 - \frac{z_{1-\alpha}}{\sqrt{n}}\sqrt{\hat
  \mu_{4,n}-(S_n^2)^2}
\end{equation}
where $z_{1-\alpha}$ is the $(1-\alpha)$-quantile of the standard
normal distribution.

This LCB can be interpreted as an estimate minus a safety margin.  By
construction, $Q_n(\alpha;\sigma^2)$ satisfies the LCB
condition~\eqref{eq:LCB_bound} with failure parameter $\alpha$, where
a truthful seller fails only with probability about $\alpha +
\oo(1/\sqrt{n})$.  It is also straightforward to check that the test
satisfies Assumption~\ref{assumption:test_tail_dens}: the sample
variance $S_n^2$ has an explicit scaled $\chi^2$ density and
concentrates at $\FishInv$, and $Q_n(\alpha;\sigma^2)$ differs from
$S_n^2$ by a vanishing $\oop(1/\sqrt{n})$ term.

Here, we compare two verification rules: (i) a ``0.05 LCB'' rule that
uses the test statistic described above with $\alpha=0.05$ and accepts
if $Q_{n_{\optagent}}(\alpha;\sigma_{\optagent}^2)\leq
\sigtil_{\optagent}^2$; and (ii) a “Sample Variance’’ rule that
accepts if the realized sample variance $S_{n_{\optagent}}^2\leq
\sigtil_{\optagent}^2$. In both cases, conditional on passing the
test, price and quantity are set as in
Mechanism~\ref{mech:cost_and_quality}, and if the test fails, the
contract is voided and the agent still incurs the acquisition cost.

\subsection{Optimal Reporting Behavior}

We examine the agent’s interim expected utility and the
corresponding optimal reported variance when the other agents report
truthfully. To isolate incentives, we fix a focal agent $\agentindex$
with opportunity cost $\cost_\agentindex=0.12$ and data variance
$\sigsq_\agentindex\in \cb{10,11,12,13,14}$.  For each verification
rule, we evaluate the focal agent’s expected utility as a function of
their reported variance $\sigtil_\agentindex^2\in[10,16]$ for three
values of the tradeoff parameter $\beta\in\cb{10,100,1000}$, and
approximate the interim expected utility using 5,000 Monte Carlo
repetitions.  Under the prior on types, the choices
$\beta\in\cb{10,100,1000}$ imply minimal procured sample sizes of
approximately 15, 50, and 158, respectively.

\begin{figure}[t]
\centering
\includegraphics[width=\linewidth]{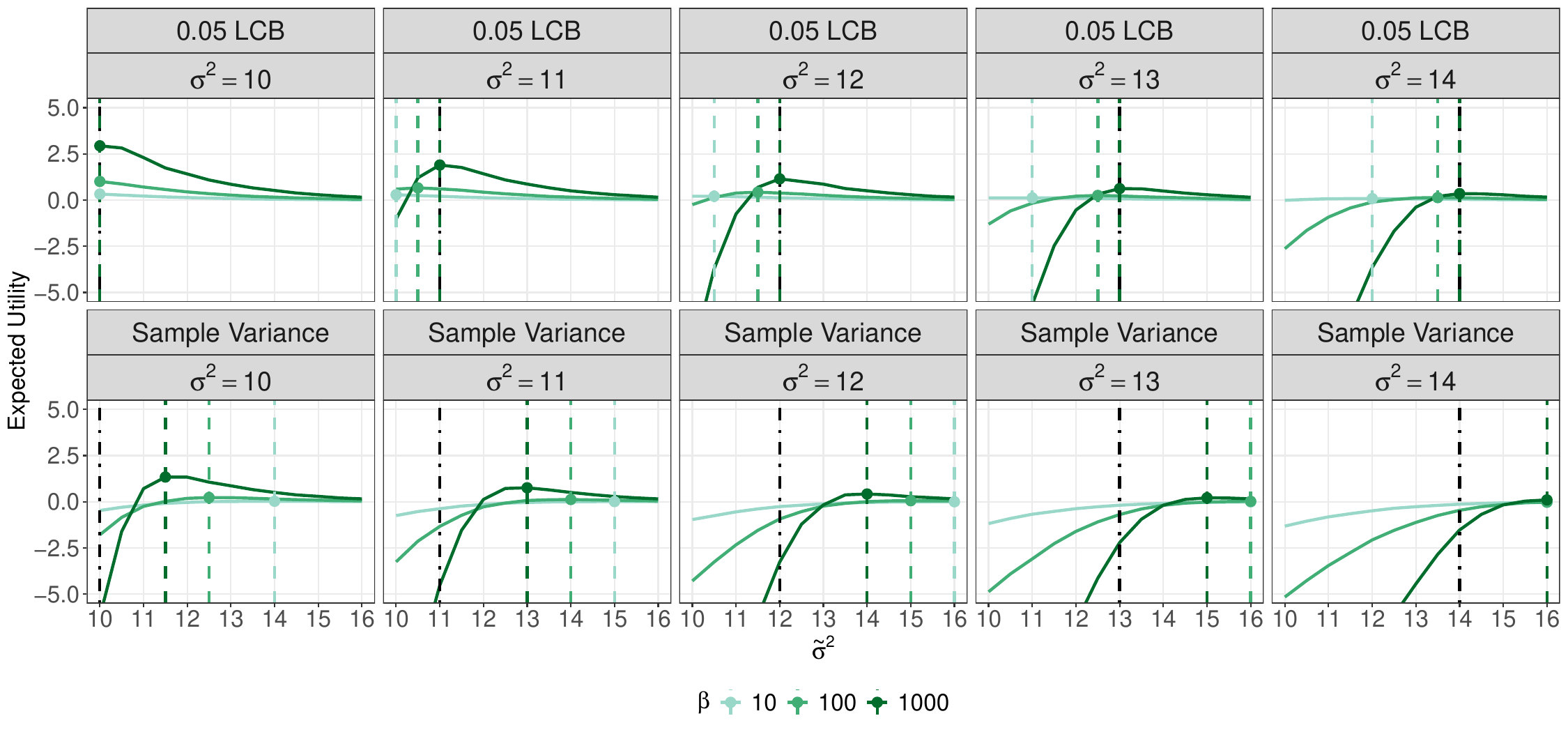}
\caption{Interim expected utility of the focal agent as a function of
  the reported variance, for different true variances (columns),
  verification rules (rows), and loss parameters (colors). The black
  dot–dashed line marks the true variance; colored dashed lines and
  dots indicate, for each $\beta$, the report that maximizes interim
  expected utility.  }
\label{fig:numerical_example_utility}
\end{figure}

\Cref{fig:numerical_example_utility} plots the resulting interim
expected utility as a function of $\sigtil_\agentindex^2$. Each panel
corresponds to a focal variance $\sigsq_\agentindex$ (columns) and a
verification rule (rows). The solid curves show
$\EE{\Util_\agentindex((\cost_\agentindex,\sigtil_\agentindex^2),
  r^{\text{tr}}_{-i}(T_{-i}); T_\agentindex)\cond
  T_\agentindex=\type_\agentindex}$ for different values of $\beta$,
where $r^{\text{tr}}$ denotes truthful reporting by the other
agents. The vertical black dot-dashed line marks the true variance
$\sigsq_\agentindex$; for each $\beta$, the colored dashed line and
filled circle indicate the report $\sigtil_\agentindex^2$ that
maximizes the focal agent’s expected utility.

Across all panels, as $\beta$ increases, the utility-maximizing report moves closer to the true variance. For moderate values of $\beta$, both under- and over-reporting can be optimal for the same agent under different verification rules: the less strict 0.05 LCB test tends to induce mild under-reporting, whereas the stricter Sample Variance test leads to more conservative over-reporting, especially for larger $\sigma^2$. 
These patterns complement our incentive analysis in~\Cref{sec:incentive_analysis}. \Cref{theorem:incentive} guarantees asymptotically optimal near-truthful reports as verification becomes increasingly informative; in this Gaussian example, the utility-maximizing quality reports themselves also move toward the truth as $\beta$ increases. The lenient LCB rule reaches this region more rapidly at finite $\beta$ than the Sample Variance rule.

\section{Proof of Theorems and Propositions}
\label{sec:proof_theo}

In this section, we collect the proofs of our main claims, including \Cref{proposition:DSIC_cost} in~\Cref{sec:mech_truth}, \Cref{proposition:DSIC_verfiable_quality} in~\Cref{sec:mech_noisy}, \Cref{theorem:incentive}, \Cref{coro:BNE} and \Cref{proposition:principal_loss} in~\Cref{sec:incentive_analysis}.


\subsection{Proof of~\Cref{proposition:DSIC_cost}}
\label{sec:proof_DSIC_cost}

We introduce the shorthand $\pricebar_\agentindex \defn \frac{\min_{j
    \neq \agentindex} s_j(\price_j)}{\FishInv_\agentindex}$.
According to the mechanism, agent $\agentindex$ wins the contract if
and only if $s_\agentindex(\price_\agentindex) = \price_\agentindex
\FishInv_\agentindex \leq \min_{j \neq \agentindex} s_j(\price_j)$, or
equivalently, if and only if $\price_\agentindex \leq
\pricebar_\agentindex$.  If agent $\agentindex$ wins, then the
associated utility is given by
\begin{equation*}
\Util_\agentindex(\price_\agentindex; \cost_\agentindex) = \p{
  \pricebar_\agentindex - \cost_\agentindex} \sqrt{\frac{\beta \,
    \FishInv_{\agentindex}}{\pricebar_\agentindex}},
\end{equation*}
which does not depend on the value $\price_\agentindex$.  On the other
hand, if $\price_\agentindex > \pricebar_\agentindex$ and agent
$\agentindex$ loses, then we have $\Util_\agentindex(\price_\agentindex;
\cost_\agentindex) = 0$ always.

Now consider the following two mutually exclusive cases:
\begin{itemize}
    \item First, suppose that $\cost_\agentindex \leq
      \pricebar_\agentindex$. Then the utility from winning the
      contract is non-negative; that is, we have
      $\Util_\agentindex(\price_\agentindex; \cost_\agentindex) \geq
      0$.  Any bid with $\price_\agentindex \leq
      \pricebar_\agentindex$ yields the same fixed non-negative
      utility, while bidding $\price_\agentindex >
      \pricebar_\agentindex$ yields a zero utility. As a consequence,
      truthfully bidding $\price_\agentindex = \cost_\agentindex \leq
      \pricebar_\agentindex$ is optimal.
    \item Otherwise, we may assume that $\cost_\agentindex >
      \pricebar_\agentindex$. In this case, if agent $\agentindex$
      were to win, then the utility from winning the contract would be
      $\Util_\agentindex(\price_\agentindex;\cost_\agentindex)< 0$, so
      that winning is strictly worse than losing. Therefore, any bid
      with $\price_\agentindex > \pricebar_\agentindex$ is optimal,
      including truthfully bidding $\price_\agentindex =
      \cost_\agentindex > \pricebar_\agentindex$.
\end{itemize}


\subsection{Proof of~\Cref{proposition:DSIC_verfiable_quality}}
\label{sec:DSIC_verfiable_quality}

We introduce the shorthand $s_{-\agentindex,(1)} = \min \big \{
s_j(\price_j, \FishInvTil_j) \, \mid j \neq \agentindex, \text{ and
  $j$ opts in} \big \}$.  According to the mechanism, agent
$\agentindex$ wins the contract if and only if
$s_\agentindex(\price_\agentindex, \FishInvTil_\agentindex) \leq
s_{-\agentindex,(1)}$, and, under exact verification, the agent passes verification if and only if $\FishInv_{\agentindex} \leq \FishInvTil_\agentindex$.
Conditional on winning, the agent's utility is
\begin{align*}
\Util_\agentindex(\price_\agentindex,\FishInvTil_\agentindex) =
I(\FishInv_{\agentindex} \leq \FishInvTil_\agentindex) \cdot
\sqrt{\beta s_{-\agentindex,(1)}} -
\frac{\sqrt{\beta}\cost_\agentindex\FishInvTil_\agentindex}{\sqrt{
    s_{-\agentindex,(1)}}}.
\end{align*}
If the agent loses, their utility is zero.

We compare the truthful report $(\cost_i, \FishInv_i)$ with an arbitrary alternative report $(\price_i, \FishInvTil_i)$. 
First suppose that $\cost_\agentindex\FishInv_\agentindex\leq
s_{-\agentindex,(1)}$.
Then the truthful report wins (up to the fixed tie-breaking rule when equality holds) and yields
\begin{align*}
\Util_\agentindex(\cost_\agentindex,\FishInv_\agentindex) =
\sqrt{\frac{\beta}{s_{-\agentindex,(1)}}} \p{ s_{-\agentindex,(1)} -
\cost_\agentindex\FishInv_\agentindex}\ge 0.
\end{align*}
In contrast, any alternative report that loses yields zero utility and therefore cannot improve upon truthful reporting. If an alternative report wins with $\FishInv_{\agentindex} > \FishInvTil_\agentindex$, it fails verification and yields strictly negative utility. Finally, if it wins with $\FishInv_{\agentindex} \le \FishInvTil_\agentindex$, then
\begin{align*}
\Util_\agentindex(\price_\agentindex,\FishInvTil_\agentindex) =
\sqrt{\frac{\beta}{s_{-\agentindex,(1)}}} \p{ s_{-\agentindex,(1)} -
\cost_\agentindex\FishInvTil_\agentindex}\le \sqrt{\frac{\beta}{s_{-\agentindex,(1)}}} \p{ s_{-\agentindex,(1)} -
\cost_\agentindex\FishInv_\agentindex},
\end{align*}
so the alternative report again cannot improve upon the truthful report.

Now suppose instead that $\cost_\agentindex\FishInv_\agentindex >
s_{-\agentindex,(1)}$, then the truthful report loses and yields zero utility. Any alternative report that also loses yields the same utility. If an alternative report wins with $\FishInv_{\agentindex} > \FishInvTil_\agentindex$, it fails verification and yields negative utility. If it wins with $\FishInv_{\agentindex} \le \FishInvTil_\agentindex$, then again
\begin{align*}
\Util_\agentindex(\price_\agentindex,\FishInvTil_\agentindex) =
\sqrt{\frac{\beta}{s_{-\agentindex,(1)}}} \p{ s_{-\agentindex,(1)} -
\cost_\agentindex\FishInvTil_\agentindex}\le \sqrt{\frac{\beta}{s_{-\agentindex,(1)}}} \p{ s_{-\agentindex,(1)} -
\cost_\agentindex\FishInv_\agentindex}\le 0.
\end{align*}
Thus, no winning deviation is profitable.


\subsection{Proof of~\Cref{theorem:incentive}}
\label{subseq:incentive}

Fix seller $\agentindex$, type $\type_\agentindex=(\cost_\agentindex,\FishInv_i)$, and a profile $a_{-\agentindex}$ of the other sellers' reports. Let $\overline s = \overline c\cdot \FishInvOver$, $\underline s = \underline c\cdot \FishInvUnder$, and $s\in [\underline s, \overline s]$ denotes the lowest competing score. If seller $\agentindex$ wins with report $a_\agentindex=(\price_\agentindex,\FishInvTil_\agentindex)$, then its expected normalized utility satisfies
\begin{equation}
\EE{\util_{\agentindex, k}((\price_\agentindex,\FishInvTil_\agentindex),
  a_{-\agentindex}; \type_\agentindex)}
 = 
\PP{\FishInvHat_\agentindex \leq  \FishInvTil_\agentindex\cond \FishInv_\agentindex,n_\agentindex} \cdot \sqrt{s} - \frac{\cost_\agentindex \FishInvTil_\agentindex}{\sqrt{ s}}.
\end{equation}
If seller $\agentindex$ loses, its utility is zero.

Let $N_k\to\infty$ be the lower sample size bound from~\Cref{lemma:sample_size_convergence}. Choose
\begin{equation}
\tau_k\defn\inf\cb{\delta>0:\zeta_{N_k}(\delta)\le \delta } +\frac{1}{N_k},\qquad
\lambda_k\defn\zeta_{N_k}(\tau_k)
\end{equation}
For every fixed $\delta>0$,~\Cref{assumption:test_tail_dens} and $N_k\to\infty$ imply $\zeta_{N_k}(\delta)\to0$. Hence, for any $\epsilon>0$ and large enough $k$,
$$\zeta_{N_k}(\epsilon/2)\le \epsilon/2
\qquad\text{and}\qquad1/N_k\le \epsilon/2,$$
which implies $\tau_k\le\epsilon$. Thus $\tau_k\to0$. Because $\zeta_{N_k}$ is nonincreasing,
$\{\delta>0:\zeta_{N_k}(\delta)\le \delta\}$ is an upper interval, and
$$\lambda_k=\zeta_{N_k}(\tau_k)\le\tau_k\to0.$$

We start by showing that reporting $\FishInvTil_\agentindex$ to be too small (i.e., substantial quality exaggeration) always leads to a non-positive expected utility uniformly over all feasible reports and is thus unprofitable.

\begin{lemma}
\label{lemma:sigma_optimal_lower}
Under the conditions of~\Cref{theorem:incentive}, for all sufficiently large $k$, every seller $\agentindex$, every type $\type_\agentindex$, every fixed profile $a_{-\agentindex}$, and every report $a_\agentindex=(\price_\agentindex,\FishInvTil_\agentindex)$ satisfying $\FishInvTil_\agentindex\le \FishInv_\agentindex - \tau_k$,
\begin{equation}
\EE{\util_{\agentindex, k}(a_\agentindex,
  a_{-\agentindex}; \type_\agentindex)} \leq 0.
\end{equation}
\end{lemma}
\noindent See~\Cref{sec:sigma_optimal_lower} for the proof.

Define the truthful information benchmark without verification
\begin{equation}
\benchmark_\agentindex(s):=\frac{\max(s-\cost_\agentindex\FishInv_i,0)}{\sqrt{s}}\geq 0.
\end{equation}
Next, we find a uniform upper bound on arbitrary reports. Intuitively, this shows that the seller's utility in this known-quality benchmark upper bounds the utility attainable by any report up to a vanishing error.

\begin{lemma}
\label{lemma:sigma_optimal_upper}
Under the conditions of~\Cref{theorem:incentive}, there exists a constant $C_1<\infty$ depending only on the bounds in~\Cref{assumption:compact} such that, for all sufficiently large $k$,
\begin{equation}
\sup_{a_i\in\mathcal A} \EE{\util_{\agentindex, k}(a_\agentindex,
  a_{-\agentindex}; \type_\agentindex)}\le 
\benchmark_\agentindex(s)+C_1\tau_k.
\end{equation}
\label{eq:u_upper_bound}
\end{lemma}
\noindent See~\Cref{sec:sigma_optimal_upper} for the proof.

Finally, consider the report in~\Cref{theorem:incentive} and its associated score under Mechanism~\ref{mech:cost_and_quality}:
\begin{equation*}
\aikDag = \left(\frac{\cost_\agentindex}{1-\lambda_k},\FishInv_\agentindex+\tau_k\right),\qquad
s_{i,k}^{\dagger}\defn \frac{\cost_\agentindex(\FishInv_\agentindex+\tau_k)}{1-\lambda_k}
\end{equation*}
We show below that (a) the report $\aikDag$ is feasible, and winning with the score $s_{i,k}^{\dagger}$ achieves a nonnegative utility, and (b) winning with the score $s_{i,k}^{\dagger}$ attains the benchmark utility $\benchmark_\agentindex(s)$ up to a vanishing loss.

\begin{lemma}
\label{lemma:global_near_optimal}
Fix an interior type $\type_\agentindex=(\cost_\agentindex,\FishInv_\agentindex)$ with $\cost_\agentindex<\overline{\cost}$, $\FishInv_\agentindex<\FishInvOver$.
For all sufficiently large $k$, $\aikDag$ is feasible, and
\begin{subequations}
\begin{equation}
\EE{\util_{\agentindex,k}(\aikDag,
a_{-\agentindex}; \type_\agentindex)}\ge0.
\label{eq:nonnegative_u}
\end{equation}
Furthermore, there exists a constant $C_2<\infty$ depending only on the bounds in~\Cref{assumption:compact} such that
\begin{equation}
\EE{\util_{\agentindex,k}(\aikDag,
a_{-\agentindex}; \type_\agentindex)}\ge \benchmark_\agentindex(s)-C_2(\tau_k+\lambda_k).
\label{eq:near_optimal_u}
\end{equation}
\end{subequations}
\end{lemma}
\noindent See~\Cref{sec:global_near_optimal} for the proof.

Putting everything together,~\Cref{eq:nonnegative_u} gives part~(a) of the claim. Combining~\Cref{eq:u_upper_bound} and~\Cref{eq:near_optimal_u} gives
\begin{equation*}
\sup_{a_\agentindex\in\mathcal A} 
\EE{\util_{\agentindex,k}(a_\agentindex,
a_{-\agentindex}; \type_\agentindex)}
- \EE{\util_{\agentindex,k}(\aikDag,
a_{-\agentindex}; \type_\agentindex)}\le (C_1+C_2)(\tau_k+\lambda_k),
\end{equation*}
and thus proves part~(b) of the claim with 
\begin{equation}
\epsilon_k\defn (C_1+C_2)(\tau_k+\lambda_k)\to 0.
\end{equation}


\subsection{Proof of~\Cref{coro:BNE}}
\label{subseq:BNE}

Fix a seller $\agentindex$ and condition on $\BigType_\agentindex=\type_\agentindex$ for $\prior_\agentindex$-almost every type $\type_\agentindex\in\TypeSet_0$. For every realized profile $a_{-\agentindex}$ of the other sellers' reports,~\Cref{theorem:incentive}(b) gives
\begin{equation*}
\sup_{a_\agentindex\in\mathcal A} 
\EE{\util_{\agentindex,k}(a_\agentindex,
a_{-\agentindex}; \type_\agentindex)}
- \EE{\util_{\agentindex,k}(\rikDag(\type_\agentindex),
a_{-\agentindex}; \type_\agentindex)}\le \epsilon_k,
\end{equation*}
where the expectation is over the randomness of the verification statistic. Moreover,~\Cref{theorem:incentive}(a) gives
\begin{equation*}
\EE{\util_{\agentindex,k}(\rikDag(\type_\agentindex),
a_{-\agentindex}; \type_\agentindex)}\ge 0
\end{equation*}
so opting out, which yields zero utility, cannot improve upon $\rikDag(\type_\agentindex)$. Thus, the bound from~\Cref{theorem:incentive}(b) continues to hold when the supremum is taken over $a_\agentindex\in\mathcal A^{\mathrm{out}}$.

Since these inequalities hold for every realized $a_{-\agentindex}$, we may take expectations over the conditional distribution of $\BigType_{-\agentindex}$ given $\BigType_{\agentindex} = \type_\agentindex$ with $a_{-\agentindex} = r_{-i,k}^{\dagger}(\BigType_{-i})$. It follows that
\begin{equation*}
\EE{\util_{\agentindex,k}(\rikDag(\type_i),
r_{-i,k}^{\dagger}(\BigType_{-i}); \type_\agentindex)\cond \BigType_\agentindex=t_\agentindex}
\ge 
\sup_{a_\agentindex\in\mathcal A^{\mathrm{out}}} 
\EE{\util_{\agentindex,k}(a_\agentindex,
r_{-i,k}^{\dagger}(\BigType_{-i}); \type_\agentindex)\cond \BigType_\agentindex=t_\agentindex}
- \epsilon_k,
\end{equation*}
and $\rkDag$ is an $\epsilon_k$-Bayesian Nash equilibrium.


\subsection{Proof of~\Cref{proposition:principal_loss}}
\label{subsec:principal_loss}

Let $\agentindex_1$, $\agentindex_2$ denote the agents with truthful scores $\score^\truth_{(1)}$ and $\score^\truth_{(2)}$. Under $\rkDag$, 
seller $\agentindex$'s reported score is
\begin{equation*}
s_{i,k}^{\dagger}=\frac{\cost_\agentindex(\FishInv_\agentindex+\tau_k)}{1-\lambda_k}
\end{equation*}
Thus, for $h=1,2$,
\begin{equation*}
s_{i_h,k}^{\dagger} \le \frac{\score^\truth_{(2)} + \cover\tau_k}{1-\lambda_k},
\end{equation*}
and therefore the second-lowest reported score satisfies
\begin{equation*}
s_{(2),k}^{\dagger} \le \frac{\score^\truth_{(2)} + \cover\tau_k}{1-\lambda_k}.
\end{equation*}
Let $\optagent$ denote the winning agent. If verification passes, the principal's loss is
\begin{equation*}
\p{1 + \frac{\FishInv_{\optagent}}{\FishInv_{\optagent}+\tau_k}}\sqrt{\beta_k
  s_{(2),k}^{\dagger}};
\end{equation*}
If verification fails, payment is voided while the delivered data are retained, so the realized loss is no larger than this quantity. Since $\FishInv_{\optagent}/\p{\FishInv_{\optagent}+\tau_k}\le 1$, 
\begin{equation*}
\loss_k \le 2\sqrt{\beta_k
  s_{(2),k}^{\dagger}}\le 2\sqrt{\beta_k
  \frac{\score^\truth_{(2)} + \cover\tau_k}{1-\lambda_k}}.
\end{equation*}
Dividing by $\loss_k^\FB = 2\sqrt{\beta_k\score^\truth_{(1)}}$ yields
\begin{equation*}
\frac{\loss_k-\loss_k^\FB}{\loss_k^\FB}
\le  \sqrt{
  \frac{\score^\truth_{(2)} + \cover\tau_k}{(1-\lambda_k)\score^\truth_{(1)}}}-1
= \sqrt{\frac{\score^\truth_{(2)}}{\score^\truth_{(1)}}}-1+ \oo\p{\tau_k+\lambda_k}.
\end{equation*}

\section{Technical Details}
\label{sec:proof_lemm}

In this section, we collect auxiliary results and proofs of lemmas, including a lower bound on the procured sample size, as well as the proofs of~\Cref{lemma:sigma_optimal_lower}--\Cref{lemma:sample_size_convergence}.

\subsection{Lower Bound on Procured Sample Size}
\label{sec:sample_size_convergence}

\begin{lemma}
\label{lemma:sample_size_convergence}
Let $j_k^*$
be the optimal provider in instance $k$, and write $n_k^* \defn
n_{j_k^*}$ for the corresponding sample size.  Under
Assumption~\ref{assumption:compact}, there exist a deterministic
sequence $\{N_k\}_{k \geq 1}$ with $N_k \to \infty$ as $k\to\infty$
such that, for all $k$ and all feasible report $(p,\FishInvTil)$ in
Mechanism~\ref{mech:cost_and_quality},
\begin{align}
n_{k}^*\ge N_k.
\end{align}
\end{lemma}
\noindent See~\Cref{sec:proof_sample_size_convergence} for the proof.


\subsection{Proof of~\Cref{lemma:sigma_optimal_lower}}
\label{sec:sigma_optimal_lower}

As the losing utility is always 0, it suffices to show that,
conditionally on winning,
\begin{equation*}
\EE{\util_{\agentindex, k}((\price_\agentindex,\FishInvTil_\agentindex),
  a_{-\agentindex}; \type_\agentindex)}
 = 
\PP{\FishInvHat_\agentindex \leq  \FishInvTil_\agentindex\cond \FishInv_\agentindex,n_\agentindex} \cdot \sqrt{s} - \frac{\cost_\agentindex \FishInvTil_\agentindex}{\sqrt{ s}}\leq 0
\end{equation*}
every report $a_\agentindex=(\price_\agentindex,\FishInvTil_\agentindex)$ satisfying $\FishInvTil_\agentindex\le \FishInv_\agentindex - \tau_k$.

By Assumption \ref{assumption:test_tail_dens},
for any $\FishInvTil_\agentindex\leq \FishInv_\agentindex-\tau_k$,
\begin{equation*}
\begin{split}
\PP{\FishInvHat_\agentindex \leq  \FishInvTil_\agentindex\cond \FishInv_\agentindex,n_\agentindex}
&\leq  \PP{\FishInvHat_\agentindex \leq  \FishInv_\agentindex-\tau_k\cond \FishInv_\agentindex,n_\agentindex}\leq  \zeta_{n_\agentindex}(\tau_k),
\end{split}
\end{equation*}
Since $n_\agentindex\ge N_k$ by~\Cref{lemma:sample_size_convergence}
\begin{equation*}
\begin{split}
\PP{\FishInvHat_\agentindex \leq  \FishInvTil_\agentindex\cond \FishInv_\agentindex,n_\agentindex}
\leq  \zeta_{N_k}(\tau_k)=\lambda_k.
\end{split}
\end{equation*}
Plugging back to the expected utility,
\begin{equation*}
\begin{split}
\EE{\util_{\agentindex, k}((\price_\agentindex,\FishInvTil_\agentindex),
  a_{-\agentindex}; \type_\agentindex)}
 &\le
\lambda_k \cdot \sqrt{s} - \frac{\cost_\agentindex \FishInvTil_\agentindex}{\sqrt{ s}}\\
&\le \frac{\lambda_k \overline s - \underline s}{\sqrt{\overline s}},
\end{split}
\end{equation*}
where $\overline s = \overline c\cdot \FishInvOver$ and $\underline s = \underline c\cdot \FishInvUnder$ are the upper and lower bounds of the scores by Assumption \ref{assumption:compact}.
Since $\lambda_k\to 0$, for all sufficiently large $k$, $\lambda_k \overline s \le \underline s$, and the right-hand side is nonpositive.


\subsection{Proof of~\Cref{lemma:sigma_optimal_upper}}
\label{sec:sigma_optimal_upper}

Consider any feasible report $a_\agentindex=(\price_\agentindex,\FishInvTil_\agentindex)$.
Recall that we defined the truthful information benchmark without verification
\begin{equation}
\benchmark_\agentindex(s):=\frac{\max(s-\cost_\agentindex\FishInv_i,0)}{\sqrt{s}}\geq 0,
\end{equation}
where $s\in [\underline s, \overline s]$ denotes the lowest competing score.

If $\FishInvTil_\agentindex\le\FishInv-\tau_k$,~\Cref{lemma:sigma_optimal_lower} gives
\begin{equation*}
\EE{\util_{\agentindex, k}(a_\agentindex,
  a_{-\agentindex}; \type_\agentindex)} \leq 0 \leq \benchmark_\agentindex(s).
\end{equation*}

Otherwise, consider the case where $\FishInvTil_\agentindex>\FishInv-\tau_k$. If the report loses, its utility is zero and the desired bound is immediate. If the report wins, since its verification-pass probability is at most one,
\begin{equation*}
\begin{split}
\EE{\util_{\agentindex, k}(a_\agentindex,
  a_{-\agentindex}; \type_\agentindex)} 
&=\PP{\FishInvHat_\agentindex \leq  \FishInvTil_\agentindex\cond \FishInv_\agentindex,n_\agentindex} \cdot \sqrt{s} - \frac{\cost_\agentindex \FishInvTil_\agentindex}{\sqrt{ s}}\\
&\leq \sqrt{s} - \frac{\cost_\agentindex \FishInvTil_\agentindex}{\sqrt{ s}}\\
&\leq \frac{s-\cost_\agentindex \FishInv_\agentindex}{\sqrt{s}} + \frac{\cost_\agentindex \tau_k}{\sqrt{ s}}\\
&\leq \benchmark_\agentindex +\frac{\overline c}{\sqrt{\underline s}}\tau_k.
\end{split}
\end{equation*}
Therefore,
\begin{equation}
\sup_{a_i\in\mathcal A} \EE{\util_{\agentindex, k}(a_\agentindex,
  a_{-\agentindex}; \type_\agentindex)}\le \benchmark_\agentindex(s)+C_1\tau_k
\end{equation}
with $C_1=\overline c/\sqrt{\underline s}$.


\subsection{Proof of~\Cref{lemma:global_near_optimal}}
\label{sec:global_near_optimal}

First, since $\type_\agentindex$ is bounded away from the upper boundary and $\lambda_k,\tau_k\to0$, the report $\aikDag$ is feasible for all sufficiently large $k$.
Furthermore, its failure probability satisfies
\begin{equation*}
\begin{split}
\PP{\FishInvHat_\agentindex >  \FishInvTil_\agentindex\cond \FishInv_\agentindex,n_\agentindex}
&\leq  \PP{\FishInvHat_\agentindex >  \FishInv_\agentindex-\tau_k\cond \FishInv_\agentindex,n_\agentindex}\\
&\leq  \zeta_{N_k}(\tau_k)=\lambda_k,
\end{split}
\end{equation*}
so its pass probability is at least $1-\lambda_k$. 
If the report loses, its utility is zero. If it wins, then $s\ge s_{i,k}^{\dagger}$, and therefore
\begin{equation*}
\begin{split}
\EE{\util_{\agentindex,k}(\aikDag,
a_{-\agentindex}; \type_\agentindex)}
&\ge (1-\lambda_k)\sqrt{s} - \frac{\cost_\agentindex(\FishInv_\agentindex+\tau_k)}{\sqrt{s}}\\
&=\frac{(1-\lambda_k)s - \cost_\agentindex(\FishInv_\agentindex+\tau_k)}{\sqrt{s}}\\
&\ge0,
\end{split}
\end{equation*}
as
\begin{equation*}
(1-\lambda_k)s \ge (1-\lambda_k)s_{i,k}^{\dagger}=\cost_\agentindex(\FishInv_\agentindex+\tau_k).
\end{equation*}
This proves~\Cref{eq:nonnegative_u}.

To prove~\Cref{eq:near_optimal_u}, consider three cases.

If $s<\cost_\agentindex\FishInv_\agentindex$, then $\benchmark_i(s)=0$.
Since $s_{i,k}^{\dagger}>\cost_\agentindex\FishInv_\agentindex$, the report $\aikDag$ loses and obtains utility zero, so the claim holds.

If $\cost_\agentindex\FishInv_\agentindex\le s\le s_{i,k}^{\dagger} $, the proposed report obtains nonnegative utility by~\Cref{eq:nonnegative_u}, which we have already shown, while
\begin{equation*}
\benchmark_\agentindex(s)=\frac{s-\cost_\agentindex\FishInv_i}{\sqrt{s}}\leq \frac{s_{i,k}^{\dagger}-\cost_\agentindex\FishInv_i}{\sqrt{\underline s}}.
\end{equation*}
For all sufficiently large $k$, e.g., when $\lambda_k\leq 1/2$,
\begin{equation*}
s_{i,k}^{\dagger}-\cost_\agentindex\FishInv_i
=\frac{\cost_\agentindex(\FishInv_\agentindex+\tau_k)}{1-\lambda_k}-\cost_\agentindex\FishInv_i\le 2\overline{c}(\tau_k+ \FishInvOver \lambda_k),
\end{equation*}
so the claim holds.

Finally, if $s> s_{i,k}^{\dagger} $, the report $\aikDag$ wins, and
\begin{equation*}
\begin{split}
\benchmark_\agentindex(s)-\EE{\util_{\agentindex,k}(\aikDag,
a_{-\agentindex}; \type_\agentindex)}
&\le 
\frac{s-\cost_\agentindex\FishInv_i}{\sqrt{s}} - \p{ (1-\lambda_k)\sqrt{s} - \frac{\cost_\agentindex(\FishInv_\agentindex+\tau_k)}{\sqrt{s}}}\\
&=\lambda_k\sqrt{s} + \frac{\cost_\agentindex\tau_k}{\sqrt{s}}\\
&\le \lambda_k\sqrt{\overline{s}} + \frac{\overline\cost\tau_k}{\sqrt{\underline s}}.
\end{split}
\end{equation*}
Combining the three cases gives~\Cref{eq:near_optimal_u} for some constant $C_2<\infty$.


\subsection{Proof of~\Cref{lemma:sample_size_convergence}}
\label{sec:proof_sample_size_convergence}

By Assumption~\ref{assumption:compact}, there exist constants
$0<\underline c\leq \bar c<\infty$ and $0<\FishInvUnder\leq
\FishInvOver<\infty$ such that every feasible report satisfies
$p\in[\underline c,\bar c]$ and
$\FishInvTil\in[\FishInvUnder,\FishInvOver]$. Thus, each score
$S_\agentindex = \price_\agentindex\FishInvTil_\agentindex$ lies in the
interval $[\underline s,\bar s]$, where $\underline s := \underline c
\FishInvUnder>0$ and $\bar s := \bar c\FishInvOver<\infty$. In
particular, for any report profile we have $S_{(2)}\leq \bar s$ and,
for the winning seller $\optagent$, $\FishInvTil_\optagent\ge
\FishInvUnder$.

Under Mechanism~2 the procured sample size for instance $k$ is
\begin{equation}
n^* = \frac{\sqrt{\beta}\FishInv_{\optagent}}{\sqrt{ S_{(2)}}} \ge
\frac{\sqrt{\beta}\FishInvUnder}{\sqrt{ \bar s}} =: N_k.
\end{equation}
This inequality holds for every $k$ and every feasible report profile.
Since $\beta_k\to\infty$ and the constants $\FishInvUnder$ and
$\bar s$ do not depend on $k$, we have $N_k\to\infty$ as $k\to\infty$.


@article{saig2023delegated,
  title={Delegated classification},
  author={Saig, Eden and Talgam-Cohen, Inbal and Rosenfeld, Nir},
  journal={Advances in Neural Information Processing Systems},
  volume={36},
  pages={13200--13236},
  year={2023}
}

@article{rochet1998ironing,
 author = {Jean-Charles Rochet and Philippe Choné},
 journal = {Econometrica},
 number = {4},
 pages = {783--826},
 publisher = {[Wiley, Econometric Society]},
 title = {Ironing, sweeping, and multidimensional screening},
 volume = {66},
 year = {1998}
}

@article{shi2025instance,
  title={Instance-adaptive hypothesis tests with heterogeneous agents},
  author={Shi, Flora C and Wainwright, Martin J and Bates, Stephen},
  journal={arXiv preprint arXiv:2510.21178},
  year={2025}
}

@article{shi2024sharp,
  title={Sharp results for hypothesis testing with risk-sensitive agents},
  author={Shi, Flora C and Bates, Stephen and Wainwright, Martin J},
  journal={arXiv preprint arXiv:2412.16452},
  year={2024}
}

@book{vaart1998asymptotic, place={Cambridge}, series={Cambridge Series in Statistical and Probabilistic Mathematics}, title={Asymptotic statistics}, publisher={Cambridge University Press}, author={van der Vaart, A. W.}, year={1998}, collection={Cambridge Series in Statistical and Probabilistic Mathematics}}

@article{ben2014optimal,
  title={Optimal allocation with costly verification},
  author={Ben-Porath, Elchanan and Dekel, Eddie and Lipman, Barton L},
  journal={American Economic Review},
  volume={104},
  number={12},
  pages={3779--3813},
  year={2014},
  publisher={American Economic Association 2014 Broadway, Suite 305, Nashville, TN 37203}
}

@article{mylovanov2017optimal,
  title={Optimal allocation with ex post verification and limited penalties},
  author={Mylovanov, Tymofiy and Zapechelnyuk, Andriy},
  journal={American Economic Review},
  volume={107},
  number={9},
  pages={2666--2694},
  year={2017},
  publisher={American Economic Association 2014 Broadway, Suite 305, Nashville, TN 37203}
}

@inproceedings{an2017towards,
  title={Towards truthful auction for big data trading},
  author={An, Dou and Yang, Qingyu and Yu, Wei and Li, Donghe and Zhang, Yang and Zhao, Wei},
  booktitle={2017 IEEE 36th International Performance Computing and Communications Conference (IPCCC)},
  pages={1--7},
  year={2017},
  organization={IEEE}
}

@article{cao2017data,
  title={Data trading with multiple owners, collectors, and users: An iterative auction mechanism},
  author={Cao, Xuanyu and Chen, Yan and Liu, KJ Ray},
  journal={IEEE Transactions on Signal and Information Processing over Networks},
  volume={3},
  number={2},
  pages={268--281},
  year={2017},
  publisher={IEEE}
}

@article{haug2011costs,
  title={The costs of poor data quality},
  author={Haug, Anders and Zachariassen, Frederik and Van Liempd, Dennis},
  journal={Journal of Industrial Engineering and Management (JIEM)},
  volume={4},
  number={2},
  pages={168--193},
  year={2011},
  publisher={Barcelona: OmniaScience}
}

@book{laffont1993theory,
  title={A theory of incentives in procurement and regulation},
  author={Laffont, Jean-Jacques and Tirole, Jean},
  year={1993},
  publisher={MIT press}
}

@article{armstrong1996multiproduct,
 author = {Mark Armstrong},
 journal = {Econometrica},
 number = {1},
 pages = {51--75},
 publisher = {[Wiley, Econometric Society]},
 title = {Multiproduct nonlinear pricing},
 volume = {64},
 year = {1996}
}

@article{yao2020price,
  title={Price-quality trade-off in procurement auctions with an uncertain quality threshold},
  author={Yao, Ying and Tanaka, Makoto},
  journal={Journal of Economic Behavior \& Organization},
  volume={177},
  pages={56--70},
  year={2020},
  publisher={Elsevier}
}

@inproceedings{chen2019prior,
  title={Prior-free data acquisition for accurate statistical estimation},
  author={Chen, Yiling and Zheng, Shuran},
  booktitle={Proceedings of the 2019 ACM Conference on Economics and Computation},
  pages={659--677},
  year={2019}
}

@article{li2021data,
  title={Data acquisition for improving machine learning models},
  author={Li, Yifan and Yu, Xiaohui and Koudas, Nick},
  journal={arXiv preprint arXiv:2105.14107},
  year={2021}
}

@inproceedings{cummings2023optimal,
  title={Optimal data acquisition with privacy-aware agents},
  author={Cummings, Rachel and Elzayn, Hadi and Pountourakis, Emmanouil and Gkatzelis, Vasilis and Ziani, Juba},
  booktitle={2023 IEEE Conference on Secure and Trustworthy Machine Learning (SaTML)},
  pages={210--224},
  year={2023},
  organization={IEEE}
}

@article{richardson2019rewarding,
  title={Rewarding high-quality data via influence functions},
  author={Richardson, Adam and Filos-Ratsikas, Aris and Faltings, Boi},
  journal={arXiv preprint arXiv:1908.11598},
  year={2019}
}

@inproceedings{ananthakrishnan2024delegating,
  title={Delegating data collection in decentralized machine learning},
  author={Ananthakrishnan, Nivasini and Bates, Stephen and Jordan, Michael and Haghtalab, Nika},
  booktitle={International Conference on Artificial Intelligence and Statistics},
  pages={478--486},
  year={2024},
  organization={PMLR}
}

@book{laffont2002theory,
  title={The theory of incentives: the principal-agent model},
  author={Laffont, Jean-Jacques and Martimort, David},
  year={2002},
  publisher={Princeton Paperbacks}
}

@article{asker2010procurement,
  title={Procurement when price and quality matter},
  author={Asker, John and Cantillon, Estelle},
  journal={The Rand journal of economics},
  volume={41},
  number={1},
  pages={1--34},
  year={2010},
  publisher={Wiley Online Library}
}

@article{huang2023evaluating,
  title={Evaluating and incentivizing diverse data contributions in collaborative learning},
  author={Huang, Baihe and Karimireddy, Sai Praneeth and Jordan, Michael I},
  journal={arXiv preprint arXiv:2306.05592},
  year={2023}
}

@article{albano2017public,
  title={Public procurement with unverifiable quality: The case for discriminatory competitive procedures},
  author={Albano, Gian Luigi and Cesi, Berardino and Iozzi, Alberto},
  journal={Journal of Public Economics},
  volume={145},
  pages={14--26},
  year={2017},
  publisher={Elsevier}
}

@article{manelli1995optimal,
  title={Optimal procurement mechanisms},
  author={Manelli, Alejandro M and Vincent, Daniel R},
  journal={Econometrica},
  volume={63},
  pages={591--620},
  year={1995},
  publisher={JSTOR}
}

@article{wan2009rfq,
  title={{RFQ} auctions with supplier qualification screening},
  author={Wan, Zhixi and Beil, Damian R},
  journal={Operations Research},
  volume={57},
  number={4},
  pages={934--949},
  year={2009},
  publisher={INFORMS}
}

@article{klein1981role,
  title={The role of market forces in assuring contractual performance},
  author={Klein, Benjamin and Leffler, Keith B},
  journal={Journal of Political Economy},
  volume={89},
  number={4},
  pages={615--641},
  year={1981},
  publisher={The University of Chicago Press}
}

@article{auriol2017economic,
  title={An economic analysis of debarment},
  author={Auriol, Emmanuelle and S{\o}reide, Tina},
  journal={International Review of Law and Economics},
  volume={50},
  pages={36--49},
  year={2017},
  publisher={Elsevier}
}

@article{board2011relational,
  title={Relational contracts and the value of loyalty},
  author={Board, Simon},
  journal={American Economic Review},
  volume={101},
  number={7},
  pages={3349--3367},
  year={2011},
  publisher={American Economic Association}
}

@article{anjarlekar2023striking,
  title={Striking a balance: An optimal mechanism design for heterogenous differentially private data acquisition for logistic regression},
  author={Anjarlekar, Ameya and Etesami, Rasoul and Srikant, R},
  journal={arXiv preprint arXiv:2309.10340},
  year={2023}
}

@article{fallah2024optimal,
  title={Optimal and differentially private data acquisition: Central and local mechanisms},
  author={Fallah, Alireza and Makhdoumi, Ali and Malekian, Azarakhsh and Ozdaglar, Asuman},
  journal={Operations Research},
  volume={72},
  number={3},
  pages={1105--1123},
  year={2024},
  publisher={INFORMS}
}

@article{jiang2023opendataval,
  title={Opendataval: a unified benchmark for data valuation},
  author={Jiang, Kevin and Liang, Weixin and Zou, James Y and Kwon, Yongchan},
  journal={Advances in Neural Information Processing Systems},
  volume={36},
  pages={28624--28647},
  year={2023}
}

@inproceedings{ghorbani2019data,
  title={Data shapley: Equitable valuation of data for machine learning},
  author={Ghorbani, Amirata and Zou, James},
  booktitle={International conference on machine learning},
  pages={2242--2251},
  year={2019},
  organization={PMLR}
}

@article{jia2019efficient,
  title={Efficient task-specific data valuation for nearest neighbor algorithms},
  author={Jia, Ruoxi and Dao, David and Wang, Boxin and Hubis, Frances Ann and Gurel, Nezihe Merve and Li, Bo and Zhang, Ce and Spanos, Costas J and Song, Dawn},
  journal={arXiv preprint arXiv:1908.08619},
  year={2019}
}

@article{li2014theory,
  title={A theory of pricing private data},
  author={Li, Chao and Li, Daniel Yang and Miklau, Gerome and Suciu, Dan},
  journal={ACM Transactions on Database Systems (TODS)},
  volume={39},
  number={4},
  pages={1--28},
  year={2014},
  publisher={Acm New York, NY, USA}
}

@article{just2023lava,
  title={Lava: Data valuation without pre-specified learning algorithms},
  author={Just, Hoang Anh and Kang, Feiyang and Wang, Jiachen T and Zeng, Yi and Ko, Myeongseob and Jin, Ming and Jia, Ruoxi},
  journal={arXiv preprint arXiv:2305.00054},
  year={2023}
}

@article{lu2024daved,
  title={DAVED: Data Acquisition via Experimental Design for Data Markets},
  author={Lu, Charles and Huang, Baihe and Karimireddy, Sai Praneeth and Vepakomma, Praneeth and Jordan, Michael and Raskar, Ramesh},
  journal={arXiv preprint arXiv:2403.13893},
  year={2024}
}

@inproceedings{wang2023data,
  title={Data banzhaf: A robust data valuation framework for machine learning},
  author={Wang, Jiachen T and Jia, Ruoxi},
  booktitle={International Conference on Artificial Intelligence and Statistics},
  pages={6388--6421},
  year={2023},
  organization={PMLR}
}

@inproceedings{yoon2020data,
  title={Data valuation using reinforcement learning},
  author={Yoon, Jinsung and Arik, Sercan and Pfister, Tomas},
  booktitle={International Conference on Machine Learning},
  pages={10842--10851},
  year={2020},
  organization={PMLR}
}

@inproceedings{zhao2023addressing,
  title={Addressing budget allocation and revenue allocation in data market environments using an adaptive sampling algorithm},
  author={Zhao, Boxin and Lyu, Boxiang and Fernandez, Raul Castro and Kolar, Mladen},
  booktitle={International Conference on Machine Learning},
  pages={42081--42097},
  year={2023},
  organization={PMLR}
}

@inproceedings{chen2011information,
  author = {Chen, Yiling and Kash, Ian A.},
  title = {Information elicitation for decision making},
  year = {2011},
  publisher = {International Foundation for Autonomous Agents and Multiagent Systems},
  address = {Richland, SC},
  booktitle = {The 10th International Conference on Autonomous Agents and Multiagent Systems - Volume 1},
  pages = {175–182},
  numpages = {8},
  series = {AAMAS '11}
}

@article{liang2018survey,
  author={Liang, Fan and Yu, Wei and An, Dou and Yang, Qingyu and Fu, Xinwen and Zhao, Wei},
  journal={IEEE Access}, 
  title={A survey on big data market: Pricing, trading and protection}, 
  year={2018},
  volume={6},
  number={},
  pages={15132-15154},
}

@inproceedings{li2022optimization,
  title={Optimization of scoring rules},
  author={Li, Yingkai and Hartline, Jason D and Shan, Liren and Wu, Yifan},
  booktitle={Proceedings of the 23rd ACM Conference on Economics and Computation},
  pages={988--989},
  year={2022}
}

@inproceedings{papireddygari2022contracts,
  title={Contracts with information acquisition, via scoring rules},
  author={Papireddygari, Maneesha and Waggoner, Bo},
  booktitle={Proceedings of the 23rd ACM Conference on Economics and Computation},
  pages={703--704},
  year={2022}
}

@inproceedings{neyman2021binary,
  title={Binary scoring rules that incentivize precision},
  author={Neyman, Eric and Noarov, Georgy and Weinberg, S Matthew},
  booktitle={Proceedings of the 22nd ACM Conference on Economics and Computation},
  pages={718--733},
  year={2021}
}

@inproceedings{oesterheld2020minimum,
  title={Minimum-regret contracts for principal-expert problems},
  author={Oesterheld, Caspar and Conitzer, Vincent},
  booktitle={International Conference on Web and Internet Economics},
  pages={430--443},
  year={2020},
  organization={Springer}
}

@article{osband1989optimal,
  title={Optimal forecasting incentives},
  author={Osband, Kent},
  journal={Journal of Political Economy},
  volume={97},
  number={5},
  pages={1091--1112},
  year={1989},
  publisher={The University of Chicago Press}
}

@article{frankel2019quantifying,
  title={Quantifying information and uncertainty},
  author={Frankel, Alexander and Kamenica, Emir},
  journal={American Economic Review},
  volume={109},
  number={10},
  pages={3650--3680},
  year={2019},
  publisher={American Economic Association 2014 Broadway, Suite 305, Nashville, TN 37203}
}

@inproceedings{lambert2008eliciting,
  title={Eliciting properties of probability distributions},
  author={Lambert, Nicolas S and Pennock, David M and Shoham, Yoav},
  booktitle={Proceedings of the 9th ACM Conference on Electronic Commerce},
  pages={129--138},
  year={2008}
}

@article{gneiting2007strictly,
  title={Strictly proper scoring rules, prediction, and estimation},
  author={Gneiting, Tilmann and Raftery, Adrian E},
  journal={Journal of the American statistical Association},
  volume={102},
  number={477},
  pages={359--378},
  year={2007},
  publisher={Taylor \& Francis}
}

@inproceedings{chen2018optimal,
  title={Optimal data acquisition for statistical estimation},
  author={Chen, Yiling and Immorlica, Nicole and Lucier, Brendan and Syrgkanis, Vasilis and Ziani, Juba},
  booktitle={Proceedings of the 2018 ACM Conference on Economics and Computation},
  pages={27--44},
  year={2018}
}

@article{castro2023data,
  title={Data-sharing markets: Model, protocol, and algorithms to incentivize the formation of data-sharing consortia},
  author={Castro Fernandez, Raul},
  journal={Proceedings of the ACM on Management of Data},
  volume={1},
  number={2},
  pages={1--25},
  year={2023},
  publisher={ACM New York, NY, USA}
}

@article{myerson1981optimal,
  title={Optimal auction design},
  author={Myerson, Roger B},
  journal={Mathematics of Operations Research},
  volume={6},
  number={1},
  pages={58--73},
  year={1981},
  publisher={INFORMS}
}

@article{miller2005eliciting,
  title={Eliciting informative feedback: The peer-prediction method},
  author={Miller, Nolan and Resnick, Paul and Zeckhauser, Richard},
  journal={Management Science},
  volume={51},
  number={9},
  pages={1359--1373},
  year={2005},
  publisher={INFORMS}
}

@article{chen2020truthful,
  title={Truthful data acquisition via peer prediction},
  author={Chen, Yiling and Shen, Yiheng and Zheng, Shuran},
  journal={Advances in Neural Information Processing Systems},
  volume={33},
  pages={18194--18204},
  year={2020}
}

@article{clarke1971multipart,
  title={Multipart pricing of public goods},
  author={Clarke, Edward H},
  journal={Public Choice},
  pages={17--33},
  year={1971},
  publisher={JSTOR}
}

@article{groves1973incentives,
  title={Incentives in teams},
  author={Groves, Theodore},
  journal={Econometrica},
  pages={617--631},
  year={1973},
  publisher={JSTOR}
}

@book{pukelsheim2006optimal,
  title={Optimal design of experiments},
  author={Pukelsheim, Friedrich},
  year={2006},
  publisher={SIAM}
}

@inproceedings{peterson2019human,
  title={Human uncertainty makes classification more robust},
  author={Peterson, Joshua C and Battleday, Ruairidh M and Griffiths, Thomas L and Russakovsky, Olga},
  booktitle={Proceedings of the IEEE/CVF international conference on computer vision},
  pages={9617--9626},
  year={2019}
}

@techreport{krizhevsky2009learning,
  author      = {Krizhevsky, Alex and Hinton, Geoffrey},
  title       = {Learning Multiple Layers of Features from Tiny Images},
  institution = {University of Toronto},
  year        = {2009},
  volume      = {1},
  number      = {4},
  pages       = {7},
  type        = {Technical Report}
}

@article{neyman1934two,
 author = {Jerzy Neyman},
 journal = {Journal of the Royal Statistical Society},
 number = {4},
 pages = {558--625},
 publisher = {[Wiley, Royal Statistical Society]},
 title = {On the two different aspects of the representative method: The method of stratified sampling and the method of purposive selection},
 volume = {97},
 year = {1934}
}

@article{vickrey1961counterspeculation,
  title={Counterspeculation, auctions, and competitive sealed tenders},
  author={Vickrey, William},
  journal={The Journal of Finance},
  volume={16},
  number={1},
  pages={8--37},
  year={1961},
  publisher={JSTOR}
}

@article{clinton2025cram,
  title={A {Cram\'{e}r--von Mises} approach to incentivizing truthful data sharing},
  author={Clinton, Alex and Zeng, Thomas and Chen, Yiding and Zhu, Xiaojin and Kandasamy, Kirthevasan},
  journal={arXiv preprint arXiv:2506.07272},
  year={2025}
}

@article{gast2020linear,
  title={Linear regression from strategic data sources},
  author={Gast, Nicolas and Ioannidis, Stratis and Loiseau, Patrick and Roussillon, Benjamin},
  journal={ACM Transactions on Economics and Computation (TEAC)},
  volume={8},
  number={2},
  pages={1--24},
  year={2020},
  publisher={ACM New York, NY, USA}
}

@inproceedings{abernethy2015low,
  title={Low-cost learning via active data procurement},
  author={Abernethy, Jacob and Chen, Yiling and Ho, Chien-Ju and Waggoner, Bo},
  booktitle={Proceedings of the Sixteenth ACM Conference on Economics and Computation},
  pages={619--636},
  year={2015}
}

@inproceedings{cai2015optimum,
  title={Optimum statistical estimation with strategic data sources},
  author={Cai, Yang and Daskalakis, Constantinos and Papadimitriou, Christos},
  booktitle={Conference on Learning Theory},
  pages={280--296},
  year={2015},
  organization={PMLR}
}

@inproceedings{ghosh2011selling,
  title={Selling privacy at auction},
  author={Ghosh, Arpita and Roth, Aaron},
  booktitle={Proceedings of the 12th ACM conference on Electronic commerce},
  pages={199--208},
  year={2011}
}

@inproceedings{fleischer2012approximately,
  title={Approximately optimal auctions for selling privacy when costs are correlated with data},
  author={Fleischer, Lisa K and Lyu, Yu-Han},
  booktitle={Proceedings of the 13th ACM conference on electronic commerce},
  pages={568--585},
  year={2012}
}
\end{document}